\documentclass{article}
\usepackage{arxiv}
\usepackage{geometry}
\usepackage{hyperref}
\usepackage{cite}

\usepackage{graphicx}
\graphicspath{{figures/}{pictures/}{images/}{./}} 

\usepackage{amsmath}
\usepackage{array}
\usepackage{microtype}
\usepackage{caption}
\usepackage{booktabs} 

\usepackage{textcomp}                  
\usepackage{mathptmx}                  
\usepackage{times}                     
\usepackage{cite}                      
\usepackage{tabu}                      
\usepackage{booktabs}                  
\usepackage[acronym]{glossaries}
\usepackage{amssymb}
\usepackage{algorithm}
\usepackage{algpseudocode}
\usepackage{enumitem}

\usepackage{graphics}
\usepackage[usenames,dvipsnames,table]{xcolor}
\usepackage{float}

\newcommand{\etal}{\textit{et~al.}}
\newcommand{\eg}{e.~g.}
\newcommand{\ie}{i.~e.}
\newcommand{\topaz}{Topaz-Denoise}

\newacronym{cryo-em}{cryo-EM}{cryogenic electron microscopy}
\newacronym{em}{EM}{Electron Microscopy}
\newacronym{stem}{STEM}{Scanning Transmission Electron Microscopy}
\newacronym{tem}{TEM}{Transmission Electron Microscopy}
\newacronym{dl}{DL}{Deep Learning}
\newacronym{gan}{GAN}{Generative Adversarial Network}
\newacronym{rl}{RL}{Reinforcement Learning}
\newacronym{vae}{VAE}{Variational Autoencoder}
\newacronym{mtf}{MTF}{Modulation Transfer Function}
\newacronym{snr}{SNR}{signal-to-noise ratio}
\newacronym{psnr}{PSNR}{peak signal-to-noise ratio}
\newacronym{ffl}{FFL}{focal frequency loss}
\newacronym{mse}{MSE}{mean square error}
\newacronym{nad}{NAD}{Nonlinear Anisotropic Diffusion}
\newacronym{nt2c}{NT2C}{Noise-Transfer2Clean}
\newacronym{pdb}{PDB}{Protein Data Bank}
\newacronym{psf}{PSF}{Point spread function}
\newacronym{difftem}{Diff-TEM}{Differentiable \gls{tem}}
\newacronym{gputem}{GPU-TEM}{GPU-\gls{tem}}
\newacronym{tmv}{TMV}{Tobacco Mosaic virus}
\newacronym{ssim}{SSIM}{structural similarity}
\newacronym{mc}{MC}{Monte Carlo}
\newacronym{dvr}{DVR}{direct volume rendering}

\DeclareMathOperator*{\argmin}{\arg\min}
\DeclareMathOperator{\poisson}{Poisson}
\DeclareMathOperator{\mtf}{MTF}
\DeclareMathOperator{\nd}{ND}
\DeclareMathOperator{\ffl}{FFL}
\DeclareMathOperator{\mse}{MSE}
\DeclareMathOperator{\normalize}{Normalize}
\DeclareMathOperator{\slf}{SL}
\DeclareMathOperator{\rd}{RD}
\DeclareMathOperator{\LL}{LL}
\DeclareMathOperator{\logl}{ll}
\DeclareMathOperator{\mean}{mean}
\DeclareMathOperator{\topazOp}{\topaz}
\DeclareMathOperator{\loss}{loss}
\DeclareMathOperator{\train}{train}

\title{Differentiable Electron Microscopy Simulation:\\Methods and Applications for Visualization}

\usepackage{authblk}
\author[1,*]{Ngan~Nguyen}
\author[1,*]{Feng~Liang}
\author[2]{Dominik~Engel}
\author[1]{Ciril~Bohak}
\author[1]{Peter~Wonka}
\author[2]{Timo~Ropinski}
\author[1]{Ivan~Viola}

\affil[1]{King Abdullah University of Science and Technology (KAUST), Saudi Arabia.
\newline
E-mails: \{ngan.nguyen $\vert$ feng.liang $\vert$ ciril.bohak $\vert$ peter.wonka $\vert$ ivan.viola\}@kaust.edu.sa.}

\affil[2]{Ulm University, Ulm, Germany.
\newline
E-mail: \{dominik.engel $\vert$ timo.ropinski\}@uni-ulm.de.}
\affil[*]{N. Nguyen and F. Liang are co-first authors.}
\date{}                     
\begin{document}
\maketitle

\begin{figure}[H]
    \centering
  \includegraphics[width=\linewidth]{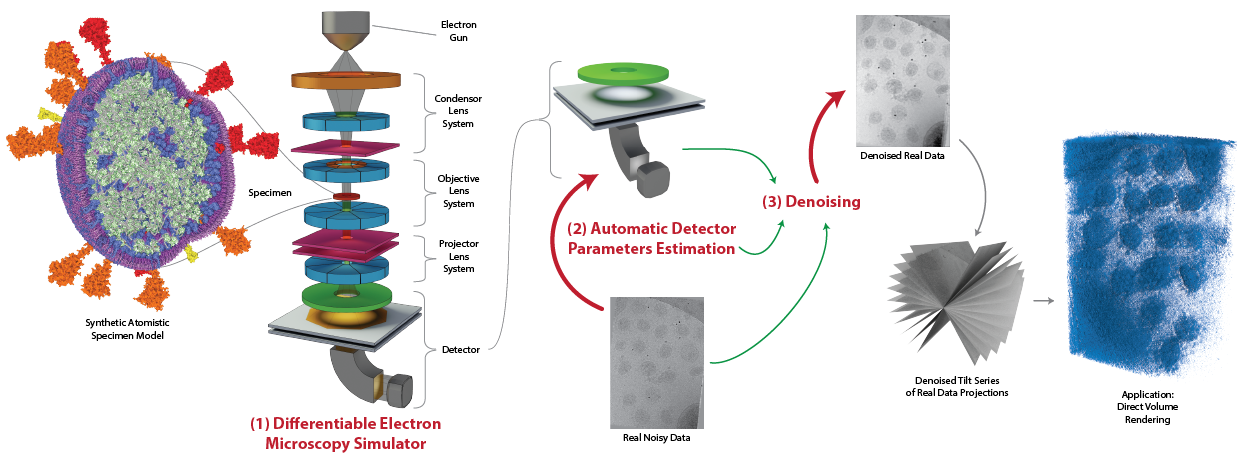}
  \caption{Three main contributions of this work are: (1) DiffTEM: a differentiable electron microscopy simulator, (2) a use case for automatic detector parameter estimation, and (3) a use case for denoising the data from real electron microscope.}
  \label{fig:teaser}
\end{figure}

\begin{abstract}
We propose a new microscopy simulation system that can depict atomistic models in a micrograph visual style, similar to results of physical electron microscopy imaging. This system is scalable, able to represent simulation of electron microscopy of tens of viral particles and synthesizes the image faster than previous methods.
On top of that, the simulator is differentiable, both its deterministic as well as stochastic stages that form signal and noise representations in the micrograph. 
This notable property has the capability for solving inverse problems by means of optimization and thus allows for generation of microscopy simulations using the parameter settings estimated from real data. We demonstrate this learning capability through two applications: (1) estimating the parameters of the modulation transfer function defining the detector properties of the simulated and real micrographs, and (2) denoising the real data based on parameters trained from the simulated examples. While current simulators do not support any parameter estimation due to their forward design, we show that the results obtained using estimated parameters are very similar to the results of real micrographs. Additionally, we evaluate the denoising capabilities of our approach and show that the results showed an improvement over state-of-the-art methods. Denoised micrographs exhibit less noise in the tilt-series tomography reconstructions, ultimately reducing the visual dominance of noise in direct volume rendering of microscopy tomograms.
\end{abstract}


\section{Introduction}
With the revolution in the resolution~\cite{kuhlbrandt2014resolution} in 2014, the \gls*{cryo-em} became an unique tool to study and visualize the cell and its inner structures at the molecular level~\cite{nogales2015cryo}. Still, extraction of the valuable information from the data obtained using \gls*{cryo-em} and developing the computational methods for processing, analyzing, and/or visualizing such data are not trivial tasks. The main reason is the fact that the data acquisition process is time-consuming, costly, and is still performed mainly by bio-experts. This situation results in a relatively small amount of data available for developing learning-centric computational methods. Additionally, due to the limitation to only use low electron beam doses to protect the specimen, the data always suffer from a low \gls*{snr}. Finally, because of the limited specimen tilting during the acquisition process, high-tilt projections are missing, which results in the missing wedge problem~\cite{Turk2020:Cryo-ET}. All these shortcomings make the usage of such data difficult for visualization. To overcome these challenges, it is possible to use highly-accurate simulated data, which are similar-to-indistinguishable to the actual acquired data. Such simulation data can be easily used in ground-truth based evaluations, since the input structural model is known prior to the microscopy simulation process. It can be used in training structural biologists to learn interpreting microscopy projections, \ie, micrographs. Such simulation can be also regarded as a distinct \emph{rendering style}, useful for example in generating animation sequences where a biological structure is transformed into a micrograph and vice versa. Last but not least, such data can be used for developing new \gls*{dl} methods for interpreting real micrographs, their testing, evaluation, and validation.

In recent years, researchers have developed several \gls*{tem} simulators to support studies in biological specimen research, some of which are based on physical principles ~\cite{rullgaard2011simulation, vulovic2013image, himes2021cryo} or parallel multi-slice simulation~\cite{lobato2015multem}.
The simulator presented by Rullg{\aa}rd~\etal~\cite{rullgaard2011simulation} uses the multi-slice method~\cite{cowley1957scattering} for modeling the interaction of electron and biological specimen, which is well suited for a computer simulation~\cite{kirkland2020advanced}. Additionally, the authors have developed a phantom generator for generating a specimen model and the calculation of its scattering potential based on a physical model. The above-mentioned simulator properties make it a good choice for simulating an acquisition process of biological specimens using \gls*{cryo-em}. However, the simulator has two major limitations. First, the simulated microscope parameters need to be calibrated manually, which is a tedious and time-consuming process where even the initial values are not available. Second, the simulator is very slow when simulating structures larger than a few proteins. Some enveloped viral particles are huge macromolecular model (\eg, ZIKA or SARS-CoV-2). It is entirely infeasible to run a microscopy simulation of many instances of such particles with any of the previous microscopy simulation systems.

In the field of differentiable rendering, visualization problems are addressed with the use of differentiable pipelines. For example, DiffDVR~\cite{Weiss2021:DiffDVR} uses a differentiable \gls*{dvr} pipeline to optimize view points, transfer functions and voxel properties. Therefore, taking inspiration from differentiable rendering, we present a fast differentiable simulator, which enables us to optimize for all continuous parameters of a microscope simulator. We demonstrate the capabilities of our differentiable simulator with two use cases, which are (1) detector parameters estimation from real data and (2) denoising. We show that using the detector parameters estimated from real data with our differentiable simulator yields accurate results. Additionally, denoising real micrographs using our pipeline is effective and outperforms state-of-the-art \gls*{dl}-based approaches.

This paper presents several technical contributions that and visualization applications:
\begin{itemize}[noitemsep,topsep=0pt]
    \item The overall architecture of the first differentiable microscopy simulator that is scalable to render realistic microscopy simulations within seconds or minutes.
    \item Learning detector parameters from real projections, to generate synthetic micrographs with the same detector characteristics that lead to similar visual appearance of the synthetic and the real projection.
    \item Utilization of deterministic and stochastic differentiation for various stages of the microscopy pipeline.
    \item Generation of large micrographs \emph{on par} with physical dimensions and complexity of real-world microscopy.
    \item A new denoising method which learns from noisy projections to create a noise-free version of real-world projections. Subsequent tomographic reconstruction then results in a noise-free volume that can be used for 3D visualization.
\end{itemize}

\section{Related work}
First we give a brief overview of \gls*{tem} simulators developed in recent years. Next, we discuss differentiable rendering and how it relates to the \gls*{tem} simulation. Then, we provide background information for gradient estimation in distribution learning. Finally, we discuss related works on \gls*{dl}-based denoising.

\paragraph{TEM Simulators:}
Several \gls*{em} simulators were developed for simulating the specimen acquisition process from different domains as presented in a book by Kirkland~\cite{kirkland2020advanced}. One of the most commonly used simulators for non-crystalline biological specimens is \gls*{tem} simulator presented by Rullg{\aa}rd \etal~\cite{rullgaard2011simulation}, which is also the basis for our differentiable version of the \gls*{tem} simulator. The presented simulator aims to provide an accurate simulation based on a well-defined physical model. Still, the implementation of the simulator introduces a set of user-adjustable parameters (\eg, detector parameters), which require a time-consuming manual calibration process. The researchers have later presented a new \gls*{tem} simulator which aims to improve the above one. While the original \gls*{tem} simulator calculates the scattering potential of the specimen by estimating the potential of the atoms, InSilico \gls*{tem}~\cite{vulovic2013image} calculates it more precisely by also including the scattering potential of molecular bonds. It also reduces the number of detector parameters and estimates them from experiments. In the recent work of Himes and Grigorieff~\cite{himes2021cryo}, they considered the \gls*{tem} simulation of amorphous samples which are sensitive to electron radiations. Additionally, they also slice the simulation in the time domain, which leads to the generation of movie frames of an exposure. The latter approaches are more accurate than the original \gls*{tem} simulator~\cite{rullgaard2011simulation} trading in additional physical aspects of the simulator with much higher computational costs. Nevertheless, all the presented simulators only support the forward simulation. To the best of our knowledge, our proposed work is the first to offer a forward and backward approach to simulation through differentiability, which is crucial in optimization for different visualization tasks.

To solve the reconstruction problem, CryoGAN~\cite{Gupta2021:CryoGAN} takes advantage of a \gls*{gan}~\cite{gan} and uses an \gls*{em} simulator based on the ASTRA Toolbox~\cite{van2015astra} for generating realistic \gls*{em} projections. During the forward calculation, it utilizes the simulation algorithm from the toolbox, while the back-projection approach from the toolbox is used to calculate gradients to enable differentiability. Given a training dataset of noisy micrographs, researchers crop empty patches that contain only background noise from the data. Next, they train a neural volume generator, whose generated volumes are fed into the simulator from the toolbox to generate clean micrographs. The empty patches of background noise are added to the clean micrographs, mocking noisy micrographs. The resulting noisy micrographs are given to the discriminator network, along with the real noisy micrographs. In contrast to our approach, this method does not model noise generation within the simulation and therefore assumes that the noise in micrographs is signal-independent. However, in the physical model of the \gls*{tem} simulator~\cite{rullgaard2011simulation}, this assumption does not hold because the models for shot noise and detector blurring depend on pixel values. Moreover, CryoGAN~\cite{Gupta2021:CryoGAN} assumes that the background noise is similar throughout the specimen, which is not the case due to the change in thickness of the specimen from center towards the boundary.

\paragraph{Differentiable Rendering:}
Together with neural rendering, it has seen dramatic progress mostly in the context of computer vision and photorealistic rendering. For example, in Li \etal~\cite{Li2018:DMC}, rendering parameters (\eg, scene geometries, camera intrinsics, and lighting conditions) can be estimated directly from images thanks to its differentiability. Hasselgren \etal~\cite{Hasselgren2020:NNTD} use a gradient estimator to optimize for ray tracing rendering parameters (\ie, sample numbers of pixels). MVTN~\cite{Hamdi2021:MVTN} takes advantage of a differentiable renderer to optimize for the performance of a downstream task (\ie, 3D geometry classification).

If we compare photorealistic rendering with \gls*{tem} simulation in a general sense, they both share similar pipelines from geometries to rendered artifacts, given a rendering method. Similar to how differentiable rendering has opened a new view on solving well-known problems in a novel way with an explainable approach, can differentiable \gls*{em} simulation provide new insight into solving simulation-related problems with an optimization approach. CryoGAN~\cite{Gupta2021:CryoGAN} already shows promising results along these lines. The main difference is that it was developed for a single-particle reconstruction and is thus lacking generalizability. In the recent work, Kniesel \etal~\cite{kniesel2022stem} developed a differentiable \gls*{stem} pipeline and showed promising results for implicit volume reconstruction. With our \gls*{tem} simulator, we can estimate \gls*{mtf} parameters of the simulator's detector part, which is in a way similar to estimating rendering parameters in a differentiable rendering method as was shown by Weiss and Westermann~\cite{Weiss2021:DiffDVR} for direct volume rendering. Moreover, our approach provides a novel approach for denoising by using images generated with a \gls*{tem} simulator in a similar way as  Noise2Noise~\cite{noise2noise} and Topaz-Denoise~\cite{topaz_denoiser} but based on a simulator physics model~\cite{rullgaard2011simulation}.

\paragraph{Gradient Estimation in Distribution Learning:}
Distribution learning has been a challenging  problem in various fields of \gls*{dl}. There are many approaches to learning a random distribution. Still, they can be classified into two categories, learning the parameters of distributions (\eg, $\mu$ and $\sigma$ of a normal distribution) and learning the manipulation of distributions, (\eg, $f_\theta(x)$ where $\theta$ parameterizes $f$, and $x$ is sampled from a prescribed distribution). 

In some scenarios, the distribution of the learning objectives of a task is not known. For example, in face generation, the distribution of faces is not known nor well-defined, which means random distributions must be prescribed and manipulated. Therefore, how to manipulate predefined random distributions is the key target of optimization. Formally, such learning can be formulated as 
\begin{equation*}
\underset{\theta}{\argmin}\ \mathbb{E}(L(f_{\theta}(X), Y)),\ X \sim D,
\end{equation*}
where $\theta$ is the set of parameters of $f$, $L$ a loss function, $D$ a prescribed distribution, and $Y$ a training dataset. \gls*{gan}~\cite{gan} is a typical case in which the optimization targets are not the parameters of the prescribed distributions but the parameters of the functions that manipulate the distributions.

When the distribution of the learning objectives is known, the distribution parameters are often the optimization target. For example, in \gls*{rl}, models can be asked to make stochastic decisions with a discrete distribution (\eg, a categorical distribution). In such settings, the optimization can be described formally as:
\begin{equation*}
\underset{\omega}{\argmin} \ \mathbb{E}(L(f(X), Y)),\ X \sim R(\omega),
\end{equation*}
where $\omega$ is the set of parameters of a distribution $R$. From the perspective of Calculus, $\frac{\partial L}{\partial \omega}$ is not defined, so researchers turn to statistical gradient estimators. One of the most widely used gradient estimators is the \gls*{mc} family. The details of \gls*{mc} gradient estimators are beyond the scope of this paper, so we refer the reader to the review on \gls*{mc} gradient estimators~\cite{Mohamed2020:MCGE}.

\gls*{vae}~\cite{Diederik2014:AEVB} is a hybrid case in which distributions' parameters and functions' parameters that manipulate the distributions are jointly learned. \gls*{vae} can be formulated as:
\begin{equation*}
\underset{\{\theta_0, \theta_1\}}{\argmin} \ \mathbb{E}(L(g_{\theta_1}(Z), Y)),\ Z \sim P(f_{\theta_0}(X)),
\end{equation*}
where $f$ and $g$ are functions paremterized by $\theta_0$ and $\theta_1$ respectively and $P$ a distribution of a prescribed type parameterized by $f_{\theta_0}(X)$.

As new and even more complex probabilistic models than \gls*{vae} emerge, existing gradient estimators and optimization techniques that are designed for specific models are not enough for training new models. Therefore, Schulman \etal~\cite{Schulman2015:GE-SCG} seek a general formulation of probabilistic models, which is stochastic compute graphs. Based on stochastic compute graphs, they devised a new way to transform deterministic loss functions into ones whose gradients are general gradient estimators of all parameters, including parameters of distributions. Such loss functions are called surrogate loss functions, and the compute graph including surrogate loss functions is called a surrogate loss function computation graph.

In our work, when estimating detector parameters, we take a \gls*{gan}-like approach, optimizing for function parameters by prescribing a distribution. In the case of denoising micrographs, we define the surrogate loss function computation graph for gradient estimation based on the work of Schulman \etal~\cite{Schulman2015:GE-SCG}.

\subsection{Denoising}
\label{denoising_related_work}
When denoising \gls*{cryo-em} micrographs, clean references are not available, so many approaches have been proposed to specifically denoise \gls*{cryo-em} micrographs, addressing this problem.

Researchers have proposed various analytical filters for denoising, such as simple averaging, Gaussian filters, and many filters in the frequency space that enhance or damp signal responses. Additionally, physics-based techniques were proposed, such as \gls*{nad}~\cite{NAD}, which takes advantage of a diffusion model to denoise 3D volumes. For a comprehensive review on filters used in \gls*{cryo-em}, we refer the reader to the review~\cite{cryo_em_filtering}.

In the era of \gls*{dl}, many neural networks are proposed to denoise \gls*{cryo-em} micrographs. Noise2Noise~\cite{noise2noise} is a general denoising approach that does not require clean data and has been adapted in CryoCARE~\cite{cryo-care} and Topaz-Denoise~\cite{topaz_denoiser}. In CryoCARE~\cite{cryo-care}, a neural network is trained as a denoiser with multiple micrographs that come from multiple consecutive exposures or multiple aligned micrographs in a tilt series or movie frames captured during a long exposure. In Topaz-Denoise~\cite{topaz_denoiser}, the researchers trained the model with a larger dataset containing thousands of micrographs acquired in different conditions, which enables better generalizability of their neural networks. Besides Noise2Noise-based approaches, \gls*{nt2c}~\cite{nt2c} trains two neural denoisers and a \gls*{gan}-based noise synthesizer to decouple noise from micrographs that are generated using a \gls*{cryo-em} simulator based on real and synthetic data.

The implicit reconstruction approach by Kniesel~\etal~\cite{kniesel2022stem} also incorporates a normalizing flows neural network to model the noise distribution in the \gls*{stem} data in a signal-dependent manner, allowing them to separate the noise from the clean reconstruction.

Our approach is similar to Noise2Noise~\cite{noise2noise} in the way that our model is required to output similar noisy images (i.e., noisy \gls*{em} projections). However, in our simulator, we do not involve neural networks but a physical model.

\section{Architecture overview}
\begin{figure}[t]
\begin{center}
\includegraphics[width=0.8\linewidth]{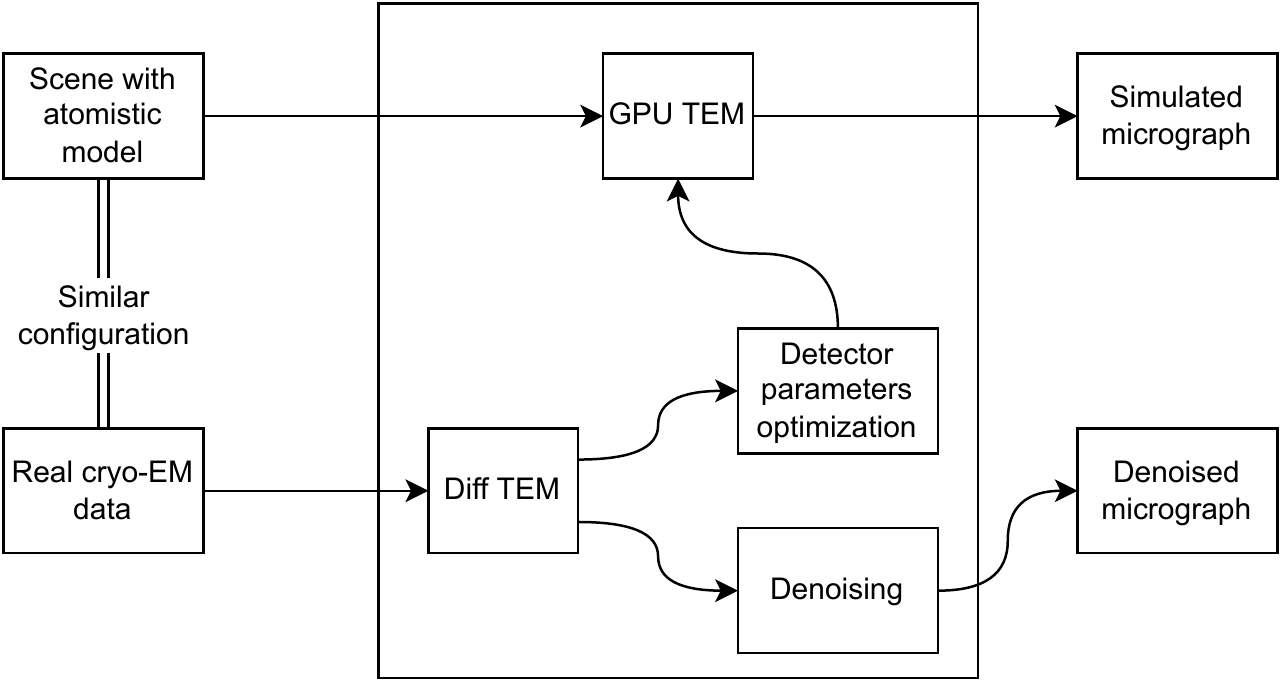}
\caption{Our \gls*{gputem} simulator receives scene configuration and detector parameters estimated from real data by \gls*{difftem} simulator for generating simulated micrograph. Our \gls*{difftem} simulator denoises the real data, the denoised data can be used for reconstruction and the reconstrution can be visualized by direct volume rendering.}
\vspace{-1.5em}
\label{fig:simulator_system}
\end{center}
\end{figure}

Our system contains two simulators as shown in Figure~\ref{fig:simulator_system}, \gls*{gputem} for rendering and \gls*{difftem} for learning. The \gls*{gputem} receives the scene with the same configuration as used for acquisition with the real \gls*{cryo-em}, and the learned parameters (\eg, \gls*{mtf}) from the \gls*{difftem} as inputs. It then simulates the synthetic data that are similar to the real data. The \gls*{difftem} receives the real \gls*{cryo-em} data and performs the learning, \eg, estimating \gls*{mtf} parameters or denoising the real data. By using this design, our system resolves two limitations of the \gls*{tem} Simulator~\cite{rullgaard2011simulation}: (1) the need for user-estimated parameters, and (2) a time-consuming simulation process for scenes containing many instances of macro-molecular structures.

\begin{figure}[t]
\begin{center}
\includegraphics[width=\linewidth]{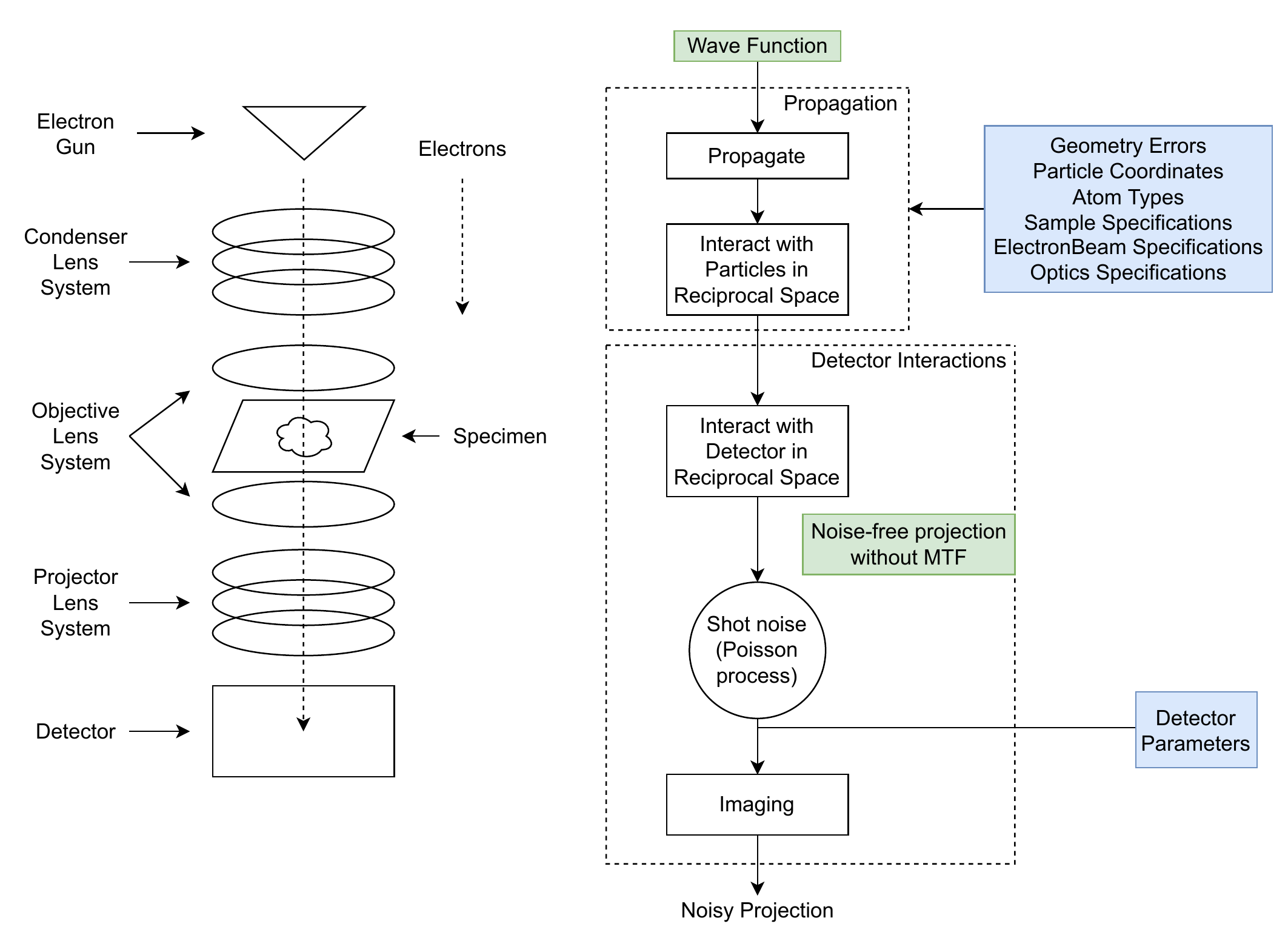}
\caption{\gls*{tem} simulator contains an electron guns, optical system with many lens, a detector and a specimen. Each electron is defined by its wave function, it interacts with specimen give the scattered wave function which is recorded at detectors to form micrographs.}
\vspace{-1.5em}
\label{fig:electron_microscope_schematic_tem}
\end{center}
\end{figure}

Both the \gls*{gputem} and the \gls*{difftem} simulators are based on the \gls*{tem} simulator~\cite{rullgaard2011simulation} and its main purpose is to simulate the image formation process. The simulator contains components similar to a real microscopy system: electron gun; specimen; optical system -- aperture, condenser lens, objective lens, projector lens, and detector. The schema of the simulator is shown in \autoref{fig:electron_microscope_schematic_tem}, where we specify the parameters with their names in blue boxes, the outputs of a step in green boxes, and stochastic nodes in circles.

To simulate the image formation process, the simulator performs two tasks. First, assembling a model of the specimen and calculating the scattering potential of the specimen are completed by the phantom generator. Second, simulating the image formation process is further split into electron-specimen interaction, electron wave propagation, and intensity detection by the detectors.

The phantom generator allows to define a specimen, including objects, background, and distribution of objects within the background. The background is assumed to be pure cryonized water. The phantom generator receives as input an atomistic model of objects defined by homologous proteins from \gls*{pdb}~\cite{Berman2003}, their distribution within the specimen (user-defined or random), sample properties (thickness, sample hole diameter), and electron gun voltage. From the inputs, it calculates the scattering potential of objects by summing the electrostatic potential of individual atoms. Next, it calculates the scattering potential of the background. Finally, it fuses background and object potentials. The details of the calculation are presented in Appendix B of the \gls*{tem} simulator paper~\cite{rullgaard2011simulation}.

Electron-specimen interaction is modeled using the Helmholtz equation presented in the work of Fanelli and {\"O}ktem~\cite{fanelli2008electron}. Electron wave propagation through the specimen is modeled using the multi-slice method~\cite{cowley1957scattering} that accounts for the thickness of the specimen and multiple scattering. The method divides the specimen into many slices with suitable thickness and describes a single electron with its wave function. For each slice, the electron wave interacts with the atoms of objects (particles) within the slice, followed by the transmission and propagation to the next slice. The process continues until the electron wave leaves the specimen.

The optical system is modeled as a convolution operator on the scattered electron wave. After leaving the specimen, the electron wave is subject to defocus, spherical aberration, and chromatic aberration of electron microscope lens. These effects are modeled through the optics contrast transfer function. 

After passing the electron microscope lens, a scattered electron wave is detected by the detector. The detector is defined as a rectangular area of square pixels and characterized by parameters such as gain, \gls*{mtf}, detective quantum efficiency, and noise sources. The primary type of noise is the shot noise~\cite{schottky1918spontane} due to fluctuations in the number of electrons which are emitted within a given time from the electron source and detected at each pixel. The electron detection in one pixel is independent with other pixels. Therefore, it is reasonably modeled with the Poisson distribution taking the pixel intensity as the expected value. In this work, we adapted the detector in line with the direct detector so other types of noise can be eliminated~\cite{levin2021direct}. The detective quantum efficiency describes the detector response of the signal. The \gls*{mtf} describes the resolution of the detector and is defined by the modulus of the Fourier transform of \gls*{psf}. The \gls*{psf} specifies the blurring of each electron contribution when detected at a pixel and propagates it into the neighboring pixels. The \gls*{psf} is a real-valued even function, so \gls*{mtf} is real valued as well. The assumption used in the \gls*{tem} simulator~\cite{rullgaard2011simulation} is that \gls*{psf} is rotationally symmetric, so we can define \gls*{mtf} as the Fourier transform of \gls*{psf}.

The \gls*{mtf} is commonly parametrized~\cite{zuo2000electron} as: 
\begin{equation}
\mtf(\omega) = \frac{a}{1+\alpha\omega^{2}} + \frac{b}{1+\beta\omega^{2}} + c,
\end{equation}
where $\omega$ is the spatial frequency in unit of 1/pixel. $a$, $b$, $c$, $\alpha > 0$, and $\beta > 0$ are the parameters depending on the type of detector and independent of the specimen.

\section{TEM Implementation}
To accelerate the \gls*{tem} simulator, we developed the \gls*{gputem} simulator that exploits parallelism of the GPU hardware using the CUDA API~\cite{cuda}. A large portion of the simulator pipeline is implemented in CUDA, while there are some  stages where we resort to the CPU implementation. We make a parallel version of the phantom generator and a per-slice parallelized version the electron-particle interaction algorithm.

For the phantom generator, the most time-consuming task is computing the scattering potential of biological structures (particles). The scattering potential of one particle is calculated by the sum of the scattering potential of all atoms that form the particle. In the original \gls*{tem}~\cite{rullgaard2011simulation} simulator, this task is performed sequentially. 
To overcome this bottleneck in our \gls*{gputem}, we make a CUDA kernel for computing scattering potential for one atom, so the scattering potential of all atoms in the particle is computed in parallel at the atom level.

To calculate the electron-specimen interaction of each slice, we first determine which structures of the specimen are within the slice. Then we calculate the scattered electron wave after it interacts with these particles. This task is also performed sequentially in the original \gls*{tem}~\cite{rullgaard2011simulation} simulator. We make another CUDA kernel in our \gls*{gputem} simulator to compute the scattered electron wave after it interacts with one structure, so the electron-specimen interaction of each slice is computed in parallel at the particle level.

To implement our \gls*{difftem} simulator, we use Pytorch~\cite{paszke2019pytorch} and Taichi~\cite{hu2019taichi, hu2019difftaichi}. Pytorch is a well-known machine learning framework for parallel computing with GPU acceleration. With Pytorch, all gradients of each component in the simulator are calculated automatically and can be used for optimization with gradient descent. However, the \gls*{tem} simulator uses the multi-slice method for computing the electron-specimen interaction, which is a sequential algorithm and hard to be expressed in Pytorch operators. Therefore, to speed up and differentiate this computation, we use Taichi~\cite{hu2019taichi, hu2019difftaichi} and implement the calculation using Taichi kernels. 

Our \gls*{difftem} simulator is differentiable, so all continuous parameters of different parts of the simulator can be optimized or learned from the real data via backpropagation and gradient descent. We demonstrate this capacity through two examples: (1) \gls*{mtf} parameters estimation and (2) denoising. From the pipeline, the final projection is obtained by adding Poisson noise to the noise-free projection and applying \gls*{mtf} afterward. Therefore, to estimate \gls*{mtf}, we perform backpropagation from real projections and use gradient descent for optimization. To denoise the real projections with gradient-based optimizations, we estimate the gradients of Poisson noise by using the method proposed by Schulman \etal~\cite{Schulman2015:GE-SCG} and apply gradient descent.

\section{Learning components}
The differentiability gives us the capability to estimate parameters along the entire electron microscopy process pipeline. In this paper we focus on detector parameters, which are at the very end of the microscopy simulation.

\subsection{MTF Parameters Estimation}

Because \gls*{mtf} parameters are independent of the specimen, we can optimize these parameters by using a few crops of background from the real projections. These crops are $M \times N$ pixels in size. They are considered as the labelled projections $I_{l_1}, I_{l_2}, \ldots, I_{l_n}$. The predicted projections $I_{p_1}, I_{p_2}, \ldots, I_{p_n}$ are created using our simulator with the same configuration for the sample, electron beam and optical system. The specimen of the predicted projection contains only the background. The predicted projections have the same size as the labelled projections. Both labelled projections and predicted projections are normalized to the range $[0, 1]$. Applying \gls*{mtf} is performed in the frequency domain which is motivation for us to choose a loss function that can capture the differences of the predicted and the labelled projection in both spatial and frequency domain. In this work, we choose \gls*{ffl}~\cite{jiang2021focal} combined with \gls*{mse} by summing them into the loss function.

\begin{figure}[tb]
\begin{center}
\includegraphics[width=\linewidth]{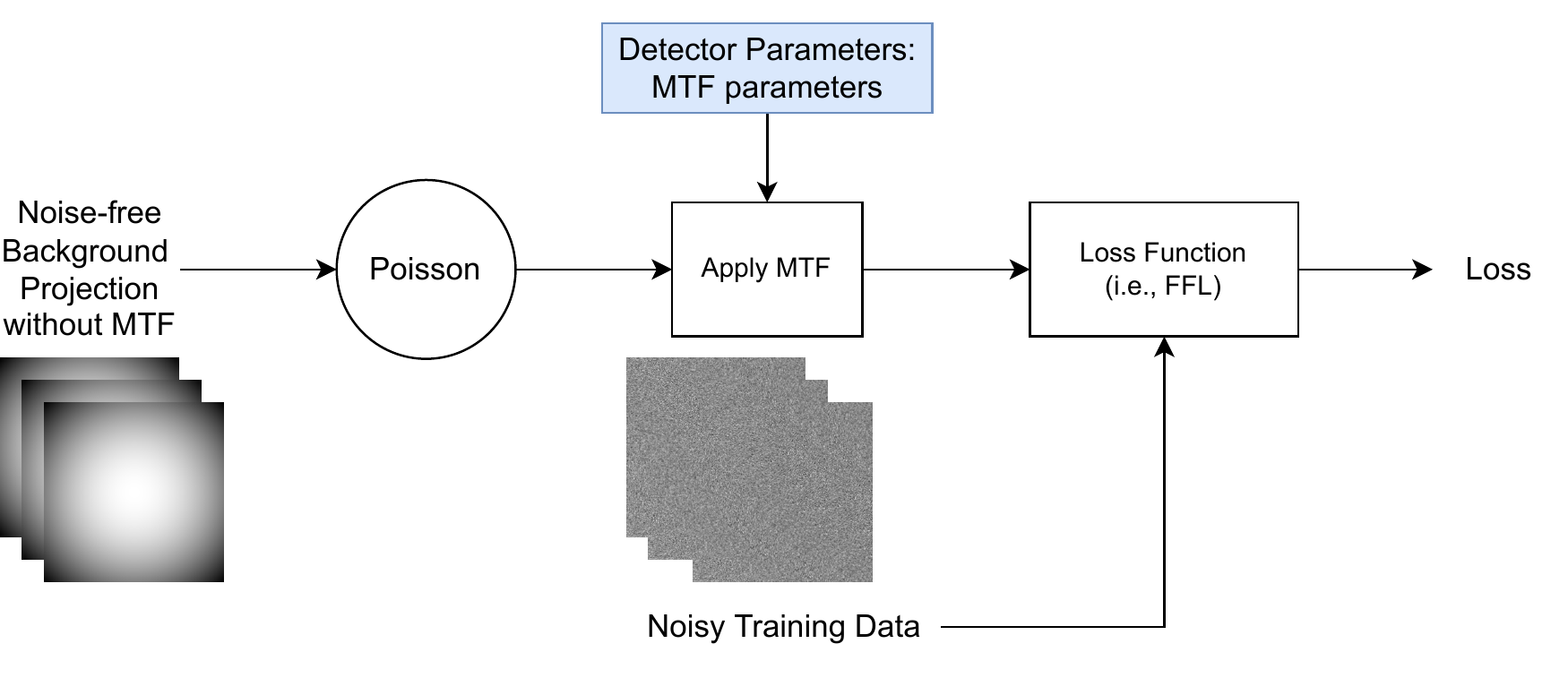}
\caption{Compute Graph of Detector Parameter Estimation: The \gls*{difftem} first receives noisy training data, computes the loss with simulated noisy data, applies the predicted \gls*{mtf}, and performs gradient descent for optimizing \gls*{mtf} parameters.}
\vspace{-1.5em}
\label{fig:compute_graph_of_detector_parameter_estimation_noisy}
\end{center}
\end{figure}

The computational graph for estimating the \gls*{mtf} parameters is shown in \autoref{fig:compute_graph_of_detector_parameter_estimation_noisy}.

The optimization can be formally described as:
\begin{equation*}
\underset{\Theta_D}{\argmin}\ \mathbb{E}(L(\mtf_{\Theta_D}(\poisson(\tilde{I})),\  \nd)),
\end{equation*}
where $\Theta_D$ is the set of \gls*{mtf} parameters, $\tilde{I}$ a noise-free projection without \gls*{mtf}, $\nd$ is the noisy dataset, and $L$ is the loss function. $\nd$ can be the synthetic or real-world data. We conducted experiments for both. 

The \gls*{ffl} loss is defined through a weight matrix and a frequency distance matrix. 
$F_{l_1}, F_{l_2}, \ldots, F_{l_n}$ are 2D discrete Fourier transforms of labelled projections $I_{l_1}, I_{l_2}, \ldots, I_{l_n}$. $F_{p_1}, F_{p_2}, \ldots, F_{p_n}$ are 2D discrete Fourier transforms of predicted projections $I_{p_1}, I_{p_2}, \ldots, I_{p_n}$.




The \gls*{ffl} of labelled projection $I_{l_i}$ and predicted projection $I_{p_i}$ is:
\begin{equation}\label{ffl}
    \ffl(F_{l_i}, F_{p_i}) = \mean(w(F_{l_i}, F_{p_i}) * \left|F_{l_i} - F_{p_i}\right|^{2})
\end{equation}
where $\left|F_{l_i} - F_{p_i}\right|$ is frequency distance between the labelled projection $I_{l_i}$ and the predicted projection $I_{p_i}$, $w(F_{l_i}, F_{p_i}) = Normalize(\left|F_{l_i} - F_{p_i}\right|^{\alpha})$ is weighted matrix, $\alpha$ is the scaling factor for flexibility. The bigger $\alpha$ is, the more focus on optimization for hard frequencies (frequency components that are hard to synthesize). We conducted an experiment with different values for $\alpha$ and chose the best one.

The \gls*{mtf} parameters are optimized using the Adam optimizer~\cite{Kingma2014} $(\beta_{1} = 0.8, \beta_{2} = 0.992, lr = 0.1)$. We also experimented with optimization using only \gls*{mse} loss. The evaluation shows that combining \gls*{ffl} and \gls*{mse} performs best for the task.

\begin{algorithm}[tb]
\small
\begin{algorithmic}
\caption{\gls*{mtf} Parameters Optimization}
\label{mtf_optimization}
\State $I_{l_n} \gets Noisy\ Projections$
\State $I_{l_n} \gets \normalize(I_{l_n})$
\State $\Theta_D \gets random\ values$
\State $\tilde{I} \gets noise\verb|-|free\ projection\ without\ \mtf$
\State $\alpha \gets Learning\ Rate$
\State $IT \gets\ iterations$
\State $i \gets 1$
\While{$i\leq IT$}
    \State $I_{p_n} \gets \mtf(\poisson(\tilde{I}))$
    \State $I_{p_n} \gets \gets \normalize(I_{p_n})$
    \State $loss \gets \mse(I_{p_n}, I_{l_n}) + \ffl(I_{p_n}, I_{l_n})$
    \State $\Theta_D\gets \Theta_D - \alpha \frac{\partial loss}{\partial \Theta_D}$
    \State $i \gets i+1$
\EndWhile
\State \Return $\Theta_D$
\end{algorithmic}
\end{algorithm}

\subsection{Denoising}
\label{denoising_section}

As shown in \autoref{fig:electron_microscope_schematic_tem}, in the forward computation of our simulator, the input of quantization is a noise-free projection without \gls*{mtf} and the output of our simulator is a noisy projection. 

Since our simulator is differentiable, we can do inverse computations via gradient descent, which means we can re-formulate the problem of denoising into an inverse problem. That is, given noisy data and a differentiable compute pipeline, to recover the noise-free projections without \gls*{mtf}, which are in the upstream of our forward computation. The optimization can be formulated as:
\begin{equation*}
\underset{\tilde{I}}{\argmin}\ \mathbb{E}(\slf(\mtf(Q(\tilde{I})),\  \rd)),
\end{equation*}
where $\tilde{I}$ is the recovered denoised projection without \gls*{mtf}, $\slf$ the surrogate loss function devised from the work of stochastic compute graph~\cite{Schulman2015:GE-SCG}, $\rd$ the real data (\eg, noisy SARS-CoV-2 projections), and $L$ a loss function. To get a denoised projection $I_d$ from $\tilde{I}$, we bypass the quantization and apply \gls*{mtf} directly, which means $I_d=\mtf(\tilde{I})$. 

\begin{figure}[tb]
\begin{center}
\includegraphics[width=\linewidth]{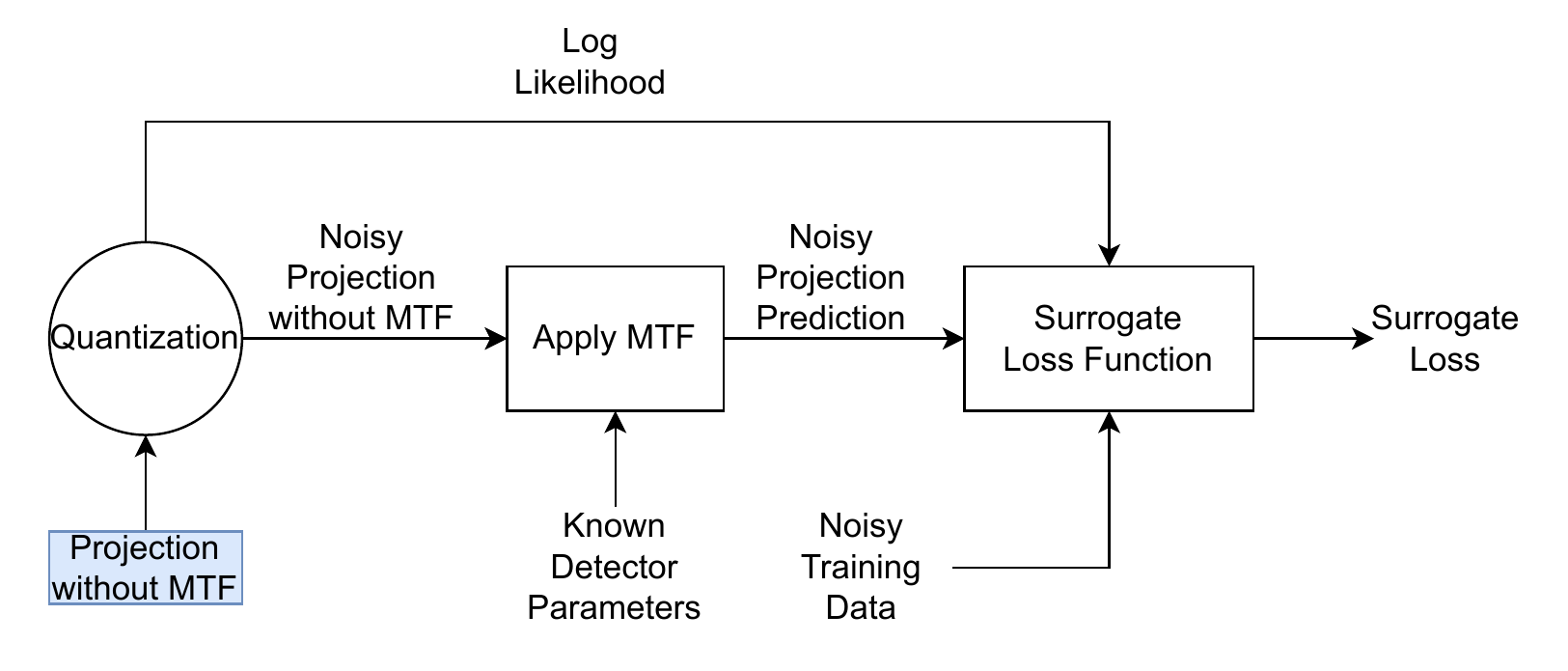}
\caption{Compute graph of denoising: The projection without \gls*{mtf} is the optimization target. Log likelihood values are used in our surrogate loss function along with the comparison loss that computes the difference between noisy projection predictions and noisy projections in a dataset.}
\vspace{-1.5em}
\label{compute_graph_of_denoising}
\end{center}
\end{figure}

The compute flow of the optimization of denoising is shown in \autoref{compute_graph_of_denoising}. The $\tilde{I}$ (\ie, \verb|Projection without MTF|) is initialized with random values. The quantization models the shot noise of a detector. Given a Poisson distribution and its parameter $\lambda$, we can calculate the log likelihood of a sampled value $x\sim \poisson(\lambda)$, which is denoted as $\LL(x,\lambda)=ln(P(x|\poisson, \lambda))$ where $ln$ is the natural logarithm. After sampling with quantization and calculating log likelihoods, the results with shot noise $I_q$ (\ie \texttt{Noisy Projection without MTF} in \autoref{compute_graph_of_denoising}) are applied with \gls*{mtf}, which yields the predicted noisy projections $I_n$.

Inspired by the surrogate loss functions by Schulman \etal~\cite{Schulman2015:GE-SCG}, we take the surrogate loss function as Eq.~\ref{surrogate_loss_denoising} where $\logl$ equals $\LL(I_q, \tilde{I})$ and $\mean(\cdot)$ is the operator taking the mean value of all values in a tensor.

\begin{equation}\label{surrogate_loss_denoising}
    \slf(I_n, \rd, \logl) = \mean((\rd-I_n)^2*\logl).
\end{equation}

The whole optimization algorithm is shown in \autoref{denoising_optimization}. The normalization trick in Line 5 ensures consistent and standardized contrast and thus stabilizes our optimization. Note that due to the scarcity of real noisy projections, we perform data augmentations to generate meaningful training data. For one micrograph $P$ in a tilt series, we align its neighboring micrographs, which are denoted as $P_{-}$ and $P_{+}$, using perspective-based image alignment in OpenCV~\cite{opencv_library} because we assume that neighboring micrographs have minor differences compared to $P$, which is a reasonable assumption when the tilt angle increment is small, for example, 2-3 degrees. Therefore, in \autoref{denoising_optimization}, $P_n = \{P_{-}, P, P_{+}\}$. Furthermore, we preprocess $P_n$ with Topaz-Denoise~\cite{topaz_denoiser} to generate training data that are less noisy. The data augmentation is optional in the synthetic setting because we have sufficient amount of noisy training data, but in the real setting, it is necessary because (1) the available raw data is scarce (\ie, only one), (2) the amount of information from only one projection with a low \gls*{snr} is limited, and (3) as discussed in Section~\ref{denoising_related_work}, our method is essentially a physical-model-based Noise2Noise~\cite{noise2noise}, which means it cannot be trained with only one projection. In the denoising optimization, we use Adam~\cite{Kingma2014} as the optimizer with $(\beta_1=0.8,\beta_2=0.992,lr=1.0)$.

\begin{algorithm}[tb]
\small
\begin{algorithmic}[1]
\caption{Denoising Optimization}
\label{denoising_optimization}
\State $P_n \gets Noisy\ Aligned\ Projections$
\State $\alpha \gets Learning\ Rate$
\State $IT \gets\ iterations$
\State $P_t \gets \topazOp(P_n)$
\State $P \gets \normalize(P_n\cup P_t)$
\State $\tilde{I} \gets random\ values$
\State $i \gets 1$
\While{$i\leq IT$}
    \State $I_q \gets Q(\tilde{I})$
    \State $\logl\gets \LL(I_q, \tilde{I})$
    \State $I_n \gets \mtf(I_q)$
    \State $\loss \gets \slf(I_n, P, ll)$
    \State $\tilde{I}\gets \tilde{I} - \alpha \frac{\partial \loss}{\partial \tilde{I}}$
    \State $i \gets i+1$
\EndWhile
\State $I_d=\mtf(\tilde{I})$
\end{algorithmic}
\end{algorithm}

\section{Reverse-Engineering MTF}
For estimating \gls*{mtf} parameters, we perform experiments on both synthetic and real-world data. The synthetic data is created from the envelope of SARS-CoV-2 and \gls*{tmv} virions. The real-world dataset is provided by Yao \etal~\cite{Sai2020:CryoET-Pipeline} and consists of 20 tilt series containing several SARS-CoV-2 virions each. For the purpose of parameter optimization we only use cropped sub-projections containing background. The real data was acquired using a Titan Krios microscope with K3 detector from Gatan, operated at a voltage of $300$ kV. The ground truth of \gls*{mtf} parameters for real data was obtained by performing curve-fitting from \gls*{mtf} curve of Gatan K3 detector\footnote{\url{https://www.gatan.com/K3\#resources}}. The ground truth values of \gls*{mtf} parameters for the synthetic data \gls*{tmv} are provided by Rullg{\aa}rd~\etal~\cite{rullgaard2011simulation} in their simulation and calibrated with their real data which is the micrograph of \gls*{tmv} virions measured by a Philips CM 200 FEG \gls*{tem} microscope.

We conducted experiments for each loss function (\gls*{mse}, \gls*{mse} + \gls*{ffl}) with varying number of background images of synthetic data to see the correlation between performance of each loss and the numbers of projections. The number that gives the best performance on synthetic data is used for the number of projections we use in experiments with real data. Furthermore, we conducted experiments with varying \gls*{ffl} loss $\alpha$ values: $(0.0, 0.5, 1.0, 2.0)$. To demonstrate the generalization of the estimation method, we also experimented with synthetic data using \gls*{mtf} varying in shape from the previous \gls*{mtf} used in synthetic and real data cases.

\begin{figure*}[t]
\begin{center}
\includegraphics[width=\textwidth]{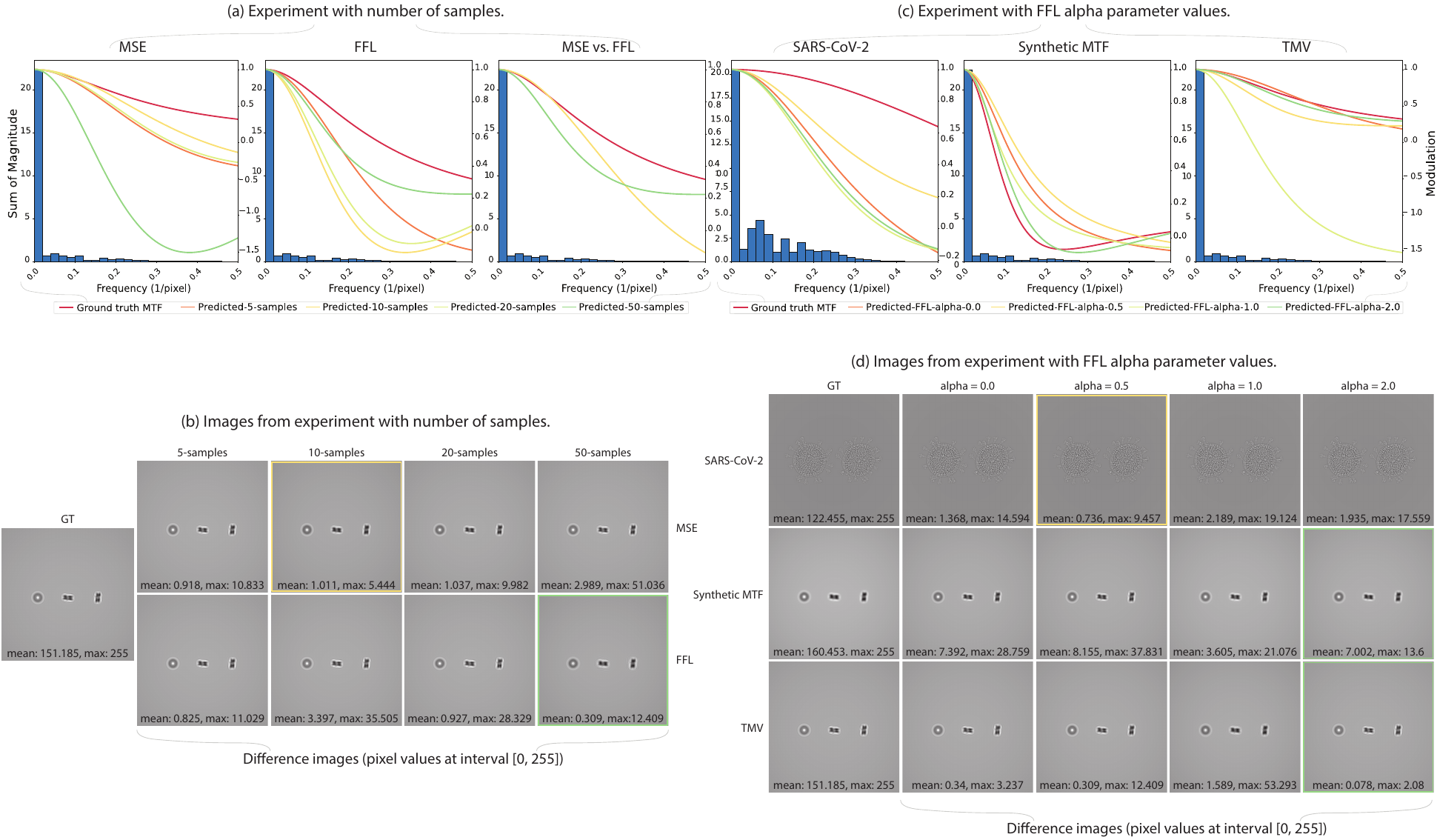}
\caption{\gls*{mtf} parameter estimation experiment results: (a) Predicted \gls*{mtf} curves estimated 
by using \gls*{mse} and \gls*{ffl} with different number of sample; (b) Noise-free projections applied \gls*{mtf}s in (a); (c) Predicted \gls*{mtf} curve for each scene using different $\alpha$ value for \gls*{ffl}; (d) Noise-free projection of each scene when applying \gls*{mtf}s in (c).}
\vspace{-1.5em}
\label{fig:mtf_parameters_exp}
\end{center}
\end{figure*}

The effect of \gls*{mtf} is difficult to visually inspect in the noisy \gls*{cryo-em} projections because of low \gls*{snr}. Therefore we applied both ground truth \gls*{mtf} and predicted \gls*{mtf} on the noise-free projection of the synthetic scene for evaluation. The details of the synthetic data, the number of background projections, and ground truth values of \gls*{mtf} parameters $(a, b, c, \alpha, \beta)$ are shown in \autoref{table:desc_mtf_exp}.

\begin{table}[t]
\small
\caption{Details of experiments for estimating \gls*{mtf} parameters}
\label{table:desc_mtf_exp}
\centering
\resizebox{0.85\linewidth}{!}{%
\begin{tabular}{lccc}
\toprule
\textsc{Subject} & \textsc{\gls*{mtf} paramters} & \textsc{No. of images}\\ 
\midrule
{3 \gls*{tmv}} & {0.7, 0.2, 0.1, 10, 40} & {5, 10, 20, 50}\\ 
{$\approx 10$ SARS-CoV-2} & {0.6, 1.24, -0.84, 0.97, 0.97} & {48} \\ 
{TMV with synthetic \gls*{mtf}} & {-2.86, 2.113, -0.1318, 57.288, 25.429} & {50} \\ 
\bottomrule
\end{tabular}
}
\end{table}

To evaluate the effect of the \gls*{mtf} on the noise-free projection, we calculated the contribution of each frequency for the noise-free projection and determined which frequency band yields the largest contribution by summing the magnitude for each band. The frequency bands (in units $\frac{1}{pixel}$) we used are $\left\{[0.0, 0.02), [0.02, 0.04), \ldots, [0.48, 0.5)\right\}$. If the predicted \gls*{mtf} aligns well with the most contributing frequency bands of the ground truth \gls*{mtf}, their effects on the noise-free projections are similar. Additionally, from the noise-free projections that applied ground truth \gls*{mtf} and predicted \gls*{mtf}, we first calculate pixel-wise difference between them, then compute the mean and maximum of the pixel-wise difference. If two \gls*{mtf}s align with the ground truth \gls*{mtf} in the same degree, the better one is the \gls*{mtf} which has the smaller mean and maximum difference values.

The correlation between the number of background projections and the performance of each loss function on synthetic data is shown in \autoref{fig:mtf_parameters_exp}~(a, b). The \gls*{mse} performance decreases when the number of background projections is greater than 10, the gap between the predicted \gls*{mtf} and the ground truth \gls*{mtf} increases, the mean and maximum differences also increase. With 50 background projections, the projections with applied predicted \gls*{mtf} are blurry, and it is hard to see the details of individual \gls*{tmv} virions. On the contrary, \gls*{ffl} performance improves when increasing the number of background projections. The predicted \gls*{mtf} with 5 background projections is more aligned with the ground truth \gls*{mtf} than 10, 20 background projections. However, the estimation from 50 background projections is better. The projection applied with the predicted \gls*{mtf} based on 50 projections has the smallest mean and maximum differences values. The comparison of the best \gls*{mse} and \gls*{ffl} results is shown in \autoref{fig:mtf_parameters_exp}~(a, b). It is clear that \gls*{ffl} is a better choice than \gls*{mse}.

We also investigated how the change of \gls*{ffl} $\alpha$ parameter affects the \gls*{mtf} estimation for different setting. The results are shown in  \autoref{fig:mtf_parameters_exp}~(c, d). The best predicted \gls*{mtf} for  SARS-CoV-2, \gls*{tmv}, and synthetic setting are obtained by using $\alpha = 0.5$, $\alpha = 2.0$, $\alpha = 2.0$, for \gls*{ffl} respectively. The best $\alpha$ value for each experiment is different because of the settings (electron dose, sample preparation, etc.) of different experiments are not the same, the distribution of frequencies in background projections is different. The results show that our system works effectively for different types of \gls*{mtf} curve. The first type is the real \gls*{mtf} (\gls*{tmv} and SARS-CoV-2), which gives high modulation in the most contributing frequency bands. This results in preservation of virion details. The second type is the synthetic \gls*{mtf}, the modulation decreases in the most contributing frequency band, resulting in blurred images with applied \gls*{mtf}.
Our predicted \gls*{mtf} for each case is aligned well with the ground truth in the most contributing frequency band, and produces similar effect on the noise-free projection.

\section{Scalable TEM Simulation}
Once we obtain the \gls*{mtf} parameters we can generate synthetic micrographs with the same detector properties as were used for a reference real micrograph. For the purpose of fast rendering we use the \gls*{gputem} implementation with several stages parallelized using CUDA.
We compare the performance of the original \gls*{tem} simulator and our \gls*{gputem}. In the phantom generator we used the model of SARS-CoV-2 virus modelled by Nguyen~\etal~\cite{nguyen2021modeling}, ZIKA virus \gls*{pdb}id 5IRE, and \gls*{tmv} PDBid 2OM3. We demonstrate the scalability of our \gls*{gputem} in two aspects: (1) number of atoms and (2) number of virion instances. The performance of each simulator for a single instance of each virus type is shown in \autoref{table:performance}. 
All performance experiments were done on the same workstation with one processor Intel(R) Xeon(R) CPU E5-2687W v4 3GHz and one GPU NVIDIA GeForece RTX 3090.

\begin{table}[t]
\small
\caption{Performance evaluation using varying viral models in terms of number of atoms and their rendering times using original simulator and \gls*{gputem}}
\label{table:performance}
\centering
\resizebox{0.75\linewidth}{!}{%
\begin{tabular}{lccccc}
\toprule
\textsc{Virions}& \textsc{No. of atoms}  & \textsc{\gls*{tem}~\cite{rullgaard2011simulation}} & \textsc{\gls*{gputem} (Ours)} \\ 
\midrule
{\gls*{tmv}} & {12505} & {45s} & \textbf{4s}\\ 
{ZIKA} & {21149} & {1m 51s} & {1m 46s} \\
{SARS-CoV-2} & {$\approx 13$ millions} & {7m 25s} & \textbf{6m 4s} \\ 
\bottomrule
\end{tabular}
}
\end{table}

From the \autoref{table:performance}, it can be seen that for a single instance of \gls*{tmv}, and a single instance of SARS-CoV-2 our \gls*{gputem} is faster than the baseline~\cite{rullgaard2011simulation}. The performance for a single instance of the ZIKA virus is similar for \gls*{gputem} and the baseline. For better understanding of simulator performance, we measure the performance on a varying number of ZIKA virion instances. The results are shown in \autoref{table:performance_instances}. Our \gls*{gputem} significantly outperforms the original implementation for scenes containing a high number of instances which are close to the realistic numbers of instances in the real microscopy experiments. Further parallelization of the simulator pipeline stages would result in an even larger performance gap.

\begin{table}[t]
\small
\caption{Performance on many instances of ZIKA}
\label{table:performance_instances}
\centering
\resizebox{0.6\linewidth}{!}{%
\begin{tabular}{rccc}
\toprule
\textsc{No. of instances} & \textsc{\gls*{tem}~\cite{rullgaard2011simulation}} & \textsc{\gls*{gputem} (Ours)}\\
\midrule
{1} &  {1m 51s} & {\textbf{1m 46s}}\\ 
{5} & {2m 9s} & {3m 1s} \\ 
{10} & {2m 16s} & {3m 10s} \\ 
{50} & {3m 34s} & {\textbf{3m 30s}} \\ 
{100} & {2m 16s} & {3m 48s} \\ 
{500} & {21m 17s} & {\textbf{5m 42s}} \\ 
{1000} & {35m 27s} & {\textbf{6m 22s}} \\ 
{5000} & {4h11m } & {\textbf{18m 9s}} \\ 
\bottomrule
\end{tabular}
}
\end{table}

\begin{figure}[tb]
\begin{center}
\includegraphics[width=\linewidth]{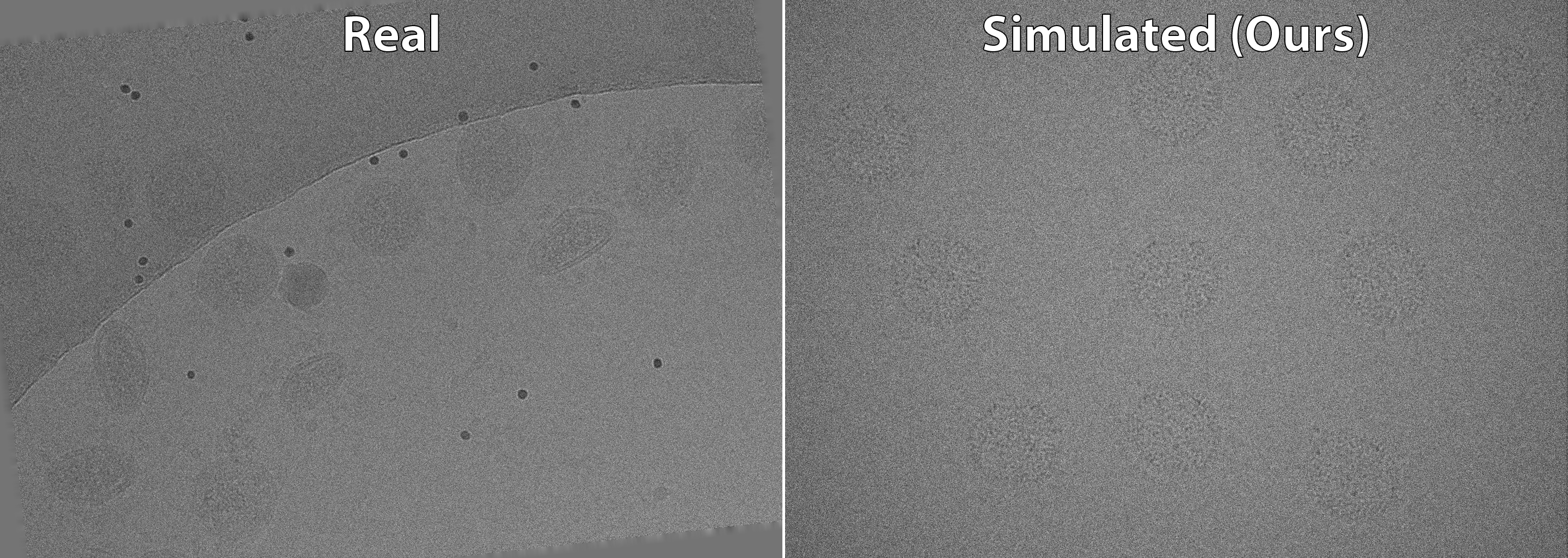}
\caption{Real and simulated tomographs of scene with multiple SARS-CoV-2 virions.}
\vspace{-1.5em}
\label{fig:many-sars-cov-virions}
\end{center}
\end{figure}

To illustrate the capability of our \gls*{gputem} simulator, in \autoref{fig:many-sars-cov-virions} we also show the result of simulating a larger scene with multiple SARS-CoV-2 virions side-by-side with the real microscope image. The presented specimens are on the same level of complexity with respect to the number of atoms/particles.

\section{Denoising Micrographs}
Our differentiable simulator \gls*{difftem} gives us the capability to predict the noise-free version of our micrographs. We have investigated this capability by means of multiple experiments, whereby we compare the denoising capabilities of our approach in synthetic and real settings qualitatively and quantitatively. 

First, we conducted experiments using a synthetic dataset generated with our simulator given \gls*{tmv} virions. In this setting, we obtain a clean ground truth data that is synthesized by bypassing the quantization. 

Note that in simulation and in the real electron microscopy process, the dose, measured in electrons per nm\textsuperscript{2} per projection, is a major factor that determines \gls*{snr}. We tested our denoising method with synthetic datasets with varying doses. The higher the dose is, the higher \gls*{snr} is. For generating a tilt series of 60 projections, 100 electrons/nm\textsuperscript{2} per projection is a typical low dose while 2000 is a typical high dose. We tested datasets acquired using doses of \{100, 200, 300, 500, 700, 1000, 1500, 2000\} per nm\textsuperscript{2} per projection. We used such doses because it is possible to use a very high dose (\eg, 1000) if the number of total projections is low and thus the total dose stays the same. Moreover, the number of available projections of the same scene (\ie, same tilt angle and specimen) also affects the denoising quality. We tested our method on the numbers of available projections of \{2, 5, 10, 25, 50\}. We denote varying data settings with pairs, \ie, (dose, number of available projections).

Moreover, to compare with the original Noise2Noise model~\cite{noise2noise} and \topaz~\cite{topaz_denoiser}, in the synthetic setting, we trained a Noise2Noise model and a \topaz~model (denoted as \texttt{Topaz T}) with our synthetic data. The data setting for training the Noise2Noise model is (100, 5000) while it is (100, 4500) for training the \topaz~model. We also use the pretrained \topaz~model (denoted as \texttt{Topaz P}) for comparison. Note that in the synthetic setting, we do not apply any kind of data augmentation in our method.

\label{sec:denoising_synth_discussion}

\begin{figure}[t]
\centering
\includegraphics[width=\linewidth]{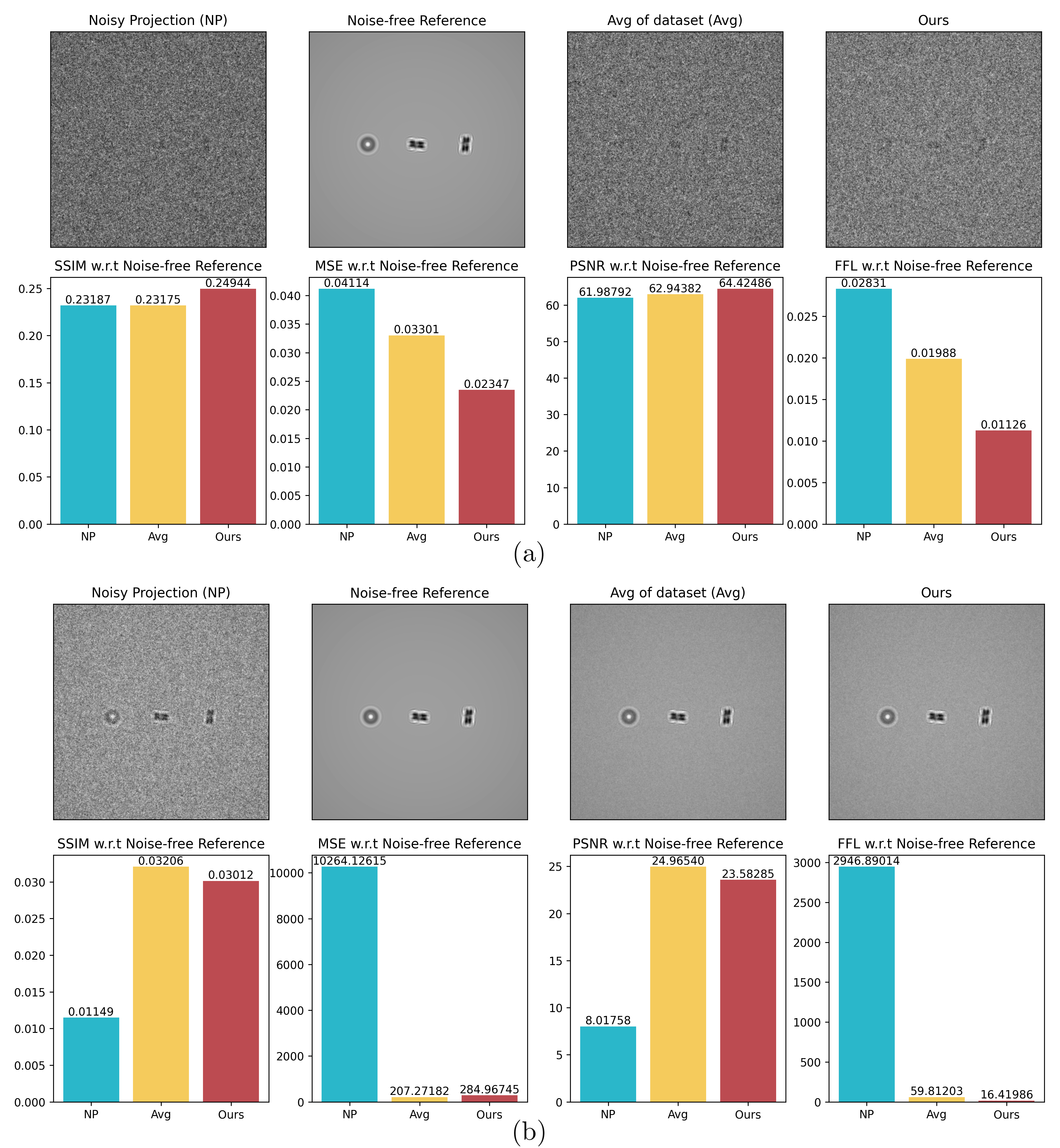}
\caption{Denoising visual and quantitative comparison of synthetic data: (a) The results of data setting (100, 2). (b) The results of data setting (2000, 50). Note that in these two settings our method (Ours) does not apply data augmentation.}
\vspace{-1.5em}
\label{denoised_results_synthetic_setting}
\end{figure}

Using the synthetic data with ground truth in the setting (100, 2) shown in \autoref{denoised_results_synthetic_setting} (a), we have very limited amount of projections (\ie, only 2) and a low dose which leads to a low \gls*{snr}. In settings like this where data are scarce and full of noise, a natural choice is to average the two available projections. Although the denoised result is not close to the noise-free reference, our method outperforms the simple average of the two available projections in terms of \gls*{ssim}~\cite{ssim}, \gls*{mse}, \gls*{psnr} and \gls*{ffl}. In the very high dose cases (\ie, {700, 1000, 1500, 2000}), we show consistently better results than simple averaging in terms of \gls*{ffl}. For example, the quantitative comparison in \autoref{denoised_results_synthetic_setting} (b) shows that the \gls*{ffl} of our method is roughly 4 times better than simple averaging. The quantitative comparisons of these two figures show that our physically-based Noise2Noise-like model is effective.

\begin{figure}[t]
\begin{center}
\includegraphics[width=\linewidth]{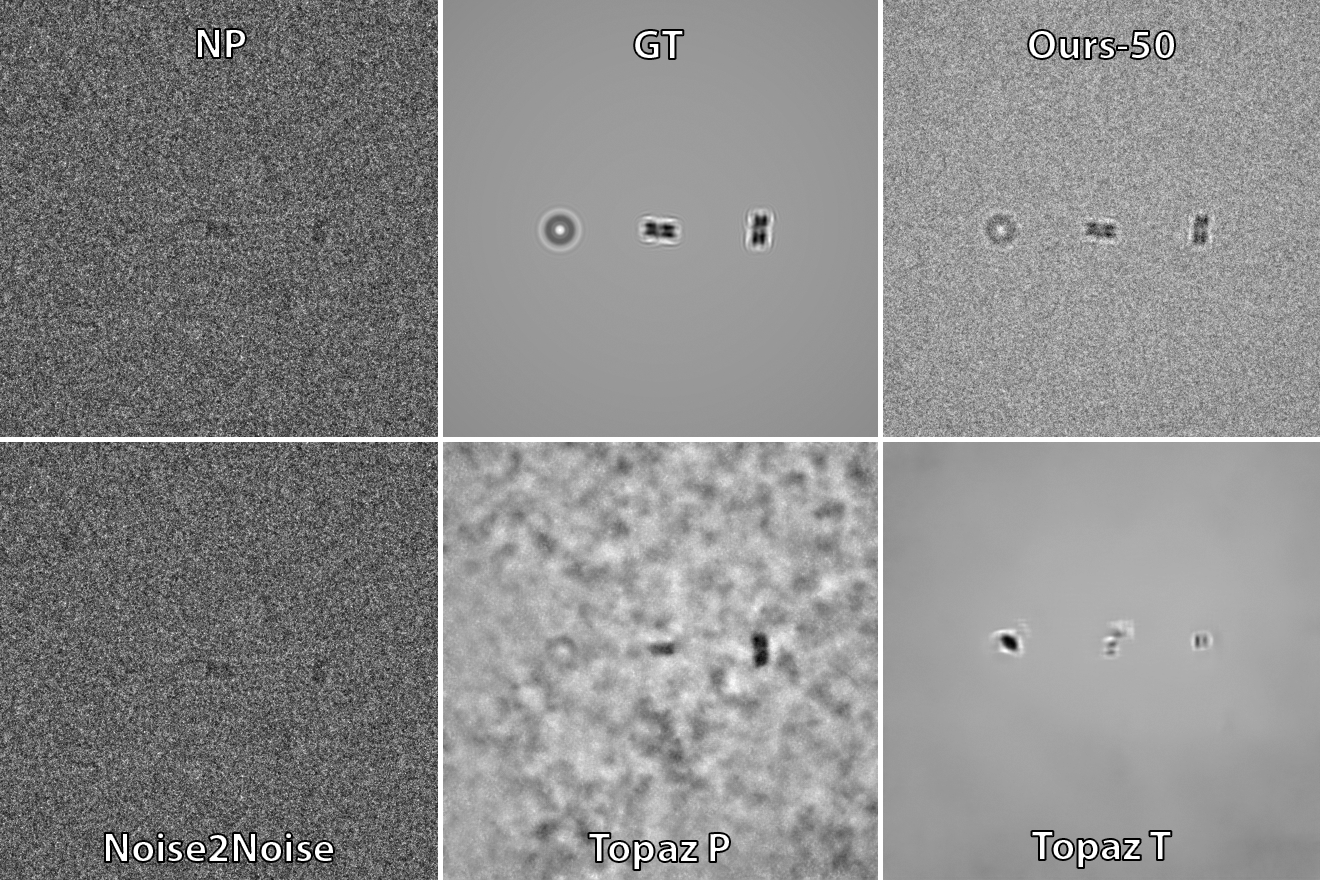}
\caption{Comparison of denoising the synthetic data: images in the top row show original noisy projection (NP) and clean ground truth projection (GT) and our denoising result, in bottom row are denoising results using the trained Noise2Noise result (Noise2Noise), pretrained \topaz~(\texttt{Topaz P}), and \topaz~trained on our noisy projections (\texttt{Topaz T}).}
\vspace{-2.5em}
\label{synthetic_denoising_comparison}
\end{center}
\end{figure}

To compare with the original Noise2Noise~\cite{noise2noise} model and \topaz~\cite{topaz_denoiser}, we present our results from the setting (100, 50) denoted as Ours-50 in \autoref{synthetic_denoising_comparison}. The figure also shows the results of a trained Noise2Noise model, \texttt{Topaz P} and \texttt{Topaz T}. The Noise2Noise model does not learn to denoise even if we have a large amount of available data (\ie, 5000). Therefore, we further tested the performance of the Topaz model with our synthetic dataset. The results show that the pretrained model (\ie, \texttt{Topaz P}) generalizes well by filtering out high-frequency noise and preserves the particles, but it is worse than our method because the boundaries of particles are blurred. Our trained Topaz model (\ie, \texttt{Topaz T}) filters out most of the noise and recognizes the positions of the particles. However, it does not preserve the shape of the particles, which makes it unusable. On the contrary, our method can work in the setting with less data (\ie, (100, 50)), which shows that our method is advantageous over training a Noise2Noise model~\cite{noise2noise} and a \topaz~model~\cite{topaz_denoiser} from scratch.

It is noteworthy that the Noise2Noise model and \texttt{Topaz T} are trained with data settings (100, 5000) and (100, 4500) respectively. It is not realistic nor viable to obtain such large datasets in real-world scenarios, which means these two models are not usable. However, \texttt{Topaz P} generalizes well in our setting, so we use it for data augmentation in the real-world scenario to generate more data for our denoising method.

For more results of our method under synthetic settings, we refer the reader to the supplementary material.

\begin{table*}[t]
\centering\caption{SSIM, PSNR and FFL comparison} \label{SSIM_PSNR_FFL_comparison}
\resizebox{\textwidth}{!}{%
\begin{tabular}{cccccc|ccccc|ccccc}
\multicolumn{6}{c}{\textbf{SSIM}} & \multicolumn{5}{c}{\textbf{PSNR}} & \multicolumn{5}{c}{\textbf{FFL}}\\
\toprule
\multicolumn{1}{c}{}      & \texttt{Noisy}    & \texttt{Topaz}    & \texttt{Ours(T)}        & \texttt{Ours(N)}         & \texttt{Ours(T+N)}      & \texttt{Noisy}    & \texttt{Topaz}    & \texttt{Ours(T)}        & \texttt{Ours(N)}        & \texttt{Ours(T+N)}      & \texttt{Noisy}    & \texttt{Topaz}    & \texttt{Ours(T)}        & \texttt{Ours(N)}   & \texttt{Ours(T+N)}      \\
\midrule
\multicolumn{1}{r|}{\texttt{Noisy}} & 1.0000   & 0.6017   & 0.4139   & 0.7395   & \textbf{0.6547}   & 361.200  & 65.8250  & 65.8270  & 70.1500  & \textbf{66.1000}  & 0.0000   & 0.0096   & 0.0082   & 0.0032   & \textbf{0.0095}   \\
\multicolumn{1}{r|}{\texttt{Topaz}} & 0.6017   & 1.0000   & 0.6894   & 0.7184   & \textbf{0.8651}   & 65.8250  & 361.200  & 72.2000  & 68.9700  & \textbf{75.0400}  & 0.0096   & 0.0000   & 0.0012   & 0.0045   & \textbf{0.0008}   \\
\multicolumn{1}{r|}{\texttt{Ours(T)}}     & 0.4139   & 0.6894   & 1.0000   & 0.7797   & 0.8001   & 65.8270  & 72.2000  & 361.200  & 71.6200  & 76.2800  & 0.0082   & 0.0012   & 0.0000   & 0.0020   & 0.0001   \\
\multicolumn{1}{r|}{\texttt{Ours(N)}}     & 0.7395   & 0.7184   & 0.7797   & 1.0000   & 0.9098   & 70.1500  & 68.9700  & 71.6200  & 361.200  & 71.6700  & 0.0032   & 0.0045   & 0.0020   & 0.0000   & 0.0029   \\
\multicolumn{1}{r|}{\texttt{Ours(T+N)}}   & 0.6547   & 0.8651   & 0.8001   & 0.9098   & 1.0000   & 66.1000  & 75.0400  & 76.2800  & 71.6700  & 361.200  & 0.0095   & 0.0008   & 0.0001   & 0.0029   & 0.0000   \\
\bottomrule
\end{tabular}
}
\end{table*}

\begin{figure*}[t]
\begin{center}
\includegraphics[width=\textwidth]{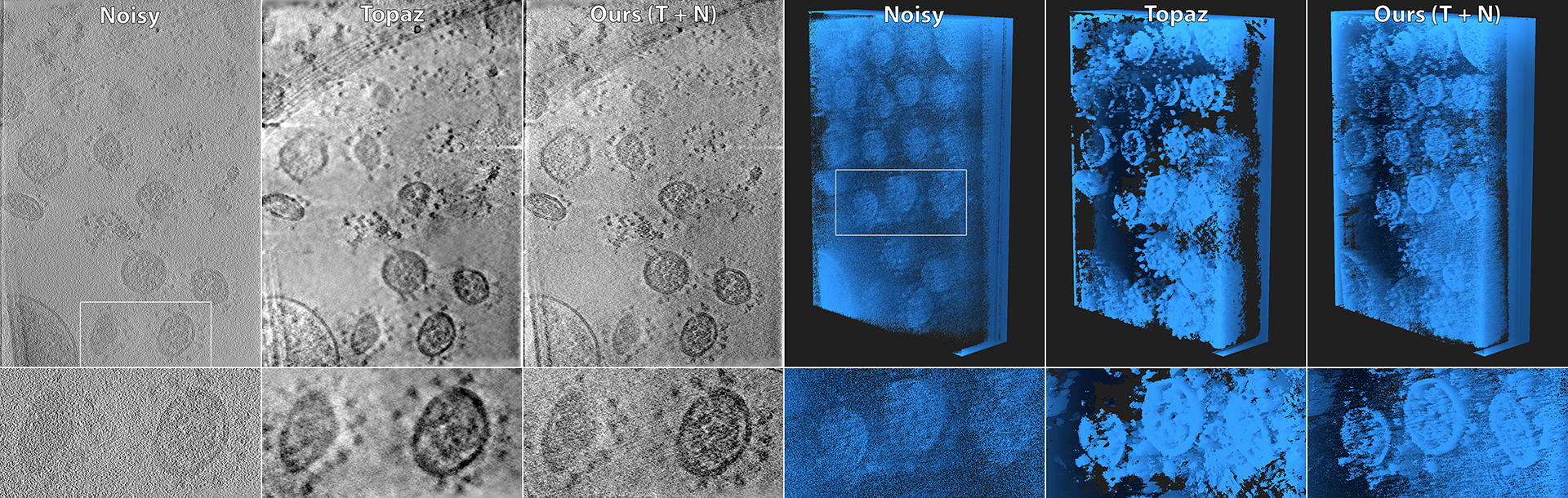}
\caption{Tomographic reconstructions from the tilt series (left) and DVR renderings (right).}
\vspace{-1em}
\label{fig:reconstruction-and-dvr-comparison}
\end{center}
\end{figure*}

In the experiments using real micrographs, we use the tilt series from Yao \etal~\cite{Sai2020:CryoET-Pipeline}, which show SARS-CoV-2 virions. The dose in the dataset is 320 electrons per nm\textsuperscript{2} per projection. Therefore, without data augmentations, the data setting is (320, 1). As discussed in Section~\ref{denoising_section}, our method cannot work in a scenario with only single projection, so in our experiments, we apply a number of data augmentations. In the setting \texttt{Ours(T)}, the data augmentation with \texttt{Topaz P} is used, yielding one more projection, which means the data setting is (320, 2). In the setting denoted as \texttt{Ours(N)}, neighboring projections of a projection are aligned and used, so the data setting is (320, 3). In the setting \texttt{Ours(T+N)}, \texttt{Topaz P} and aligned neighboring projections are used, so the data setting is (320, 6). The case of \texttt{Ours(T+N)} matches exactly our \autoref{denoising_optimization} while in the case of \texttt{Ours(T)}, $P_n=\{P\}$, and in the case of \texttt{Ours(N)}, preprocessing with \topaz~is not used (\ie, $P_{\train}=\normalize(P_n)$). In the real setting, we do not have noise-free ground truth projections to compare against, so we first present results of the qualitative comparison.

\begin{figure}[t]
\begin{center}
\includegraphics[width=\linewidth]{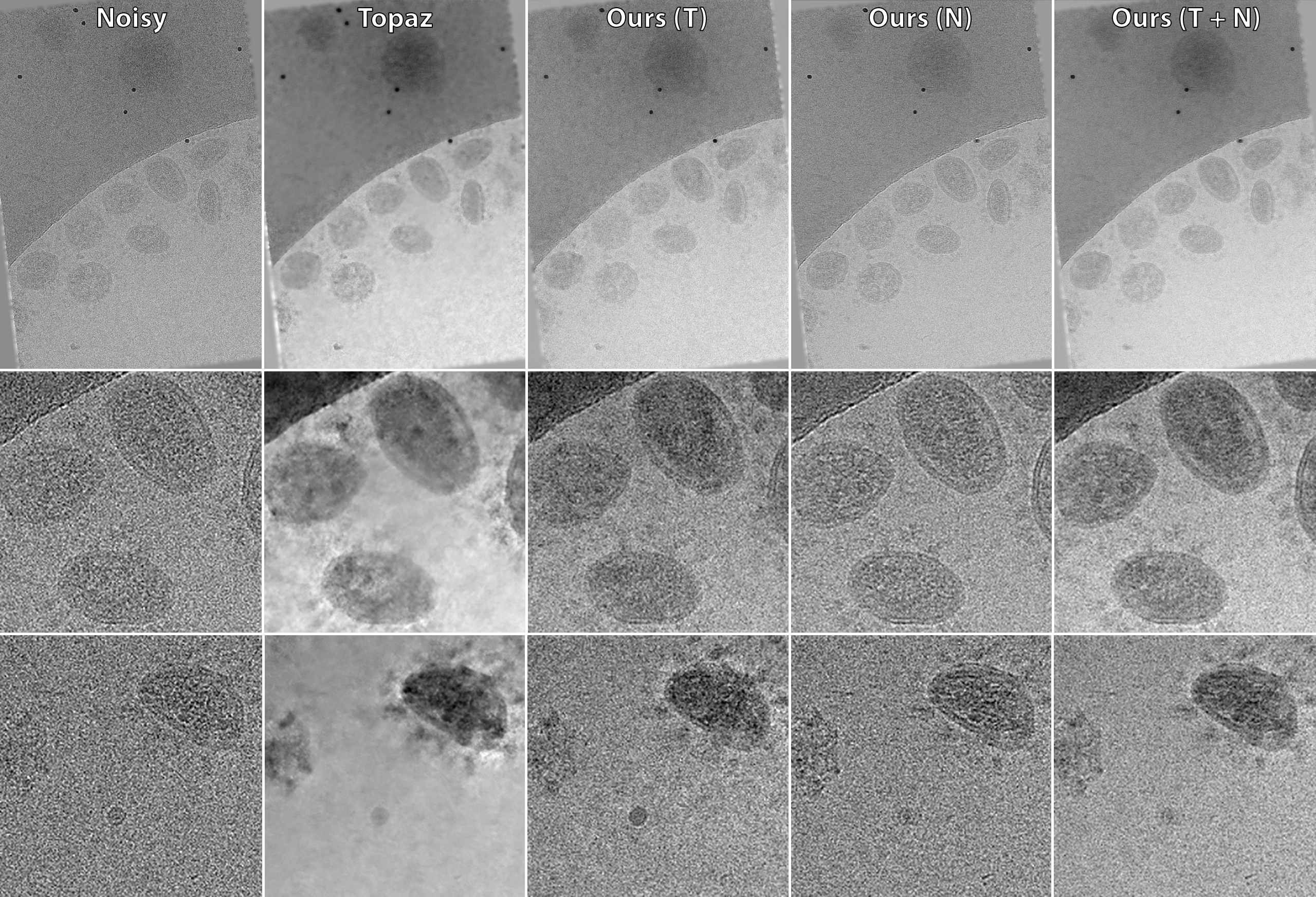}
\caption{\texttt{Topaz}: the real projection denoised by \texttt{Topaz P}; \texttt{Noisy}: the real noisy projection; Top: Full views; Middle: Zoom-in of the first row. In the bottom row is another example of our denoising method, \texttt{Ours(T+N)} contains less noise while preserving boundaries.}
\vspace{-2em}
\label{denoising_with_real_data}
\end{center}
\end{figure}

On the top of \autoref{denoising_with_real_data}, we can see \texttt{Topaz} has block-like artifacts and \texttt{Noisy} contains more noise while others have minor visual difference. However, if we zoom-in the images, as shown in the bottom of ~\autoref{denoising_with_real_data}, we can see (1) \texttt{Topaz} has more blurring artifacts, (2) \texttt{Noisy} has the most noise, and (3) \texttt{Ours(T+N)} is better than \texttt{Ours(N)} and \texttt{Ours(T)} in that it has less high-frequency noise while preserving clear boundaries. This makes our method \texttt{Ours(T+N)} the best among these five because of its good balance between denoising and preserving boundaries from the view of quality. 

We also present quantitative results in Table~\ref{SSIM_PSNR_FFL_comparison}. Based on the bottom images from \autoref{denoising_with_real_data}, we can conclude that \texttt{Topaz} contains the least noise while \texttt{Noisy} contains the most noise. With this conclusion, the values in bold in the table show that (1) \texttt{Ours(T+N)} is the closest one to \texttt{Topaz} and (2) \texttt{Ours(T+N)} has a moderate amount of noise compared to \texttt{Ours(T)} and \texttt{Ours(N)}. This means \texttt{Ours(T+N)} is a good compromise between removing noise and preserving details, echoing with our qualitative comparison.

In addition to the example on the top two rows of \autoref{denoising_with_real_data}, we give another example of denoising real data. From the last row in \autoref{denoising_with_real_data}, we can draw similar conclusions. \texttt{Topaz} has the least noise, but compromised boundaries while \texttt{Ours(T+N)} has a moderate amount of noise but also clearer boundaries and more visibility of spikes.

The application of such denoising are also useful for the tomographic reconstruction. In left part of  \autoref{fig:reconstruction-and-dvr-comparison} we show the mid-slice from the volumes reconstructed using SIRT reconstruction technique from IMOD~\cite{kremer1996computer} from original noisy projections (left), from denoised projections using \texttt{Topaz} (middle), and from denoised projections using our approach \texttt{Ours(T+N)} (right). One can clearly see that \texttt{Topaz} and our approach are better that reconstruction from the original noisy projections. Moreover, our approach retains more details which are smudged out by \texttt{Topaz}.

Additionally, in the right part of \autoref{fig:reconstruction-and-dvr-comparison}, we show \gls*{dvr} rendering of the same volumes using the same rendering parameters apart from the threshold, which was manually adapted for each case. One can see that \texttt{Topaz} cleans the volume extensively, but also removes parts of the virions and removes the high frequency details from the structures. This is not the case for results using our approach. While there is more noise present in the volume, the high-frequency details are retained. Both approaches are clearly better choice than rendering the original noisy volume.

\section{Conclusion}
With this work, we have presented a system combining \gls*{gputem} and \gls*{difftem} simulators that enables rendering large micrographs \emph{on par} with physical dimensions and complexity of real-world microscopy. Nonetheless, our system not only performs forward calculation, but also enables backward computation, which gives rise to a new way to solve inverse problems in microscopy imaging. We demonstrate its ability by showcasing two examples, detector parameter estimation, and denoising, which are typical inverse problems.

Our \gls*{gputem} simulator outperforms the baseline \gls*{tem} simulator when simulating a scene containing a high number of complex structures. However, because the calculation for all slices of the specimen is still performed on the CPU, our \gls*{gputem} still takes time to transfer the data from GPU to CPU for this calculation. In the future, we plan to perform this calculation in parallel.

Our method for detector parameters estimation learns from the real data, automating the time-consuming manual calibration. Moreover, users can use this functionality to guess a reasonable set of parameters and fine-tune them  manually if desired. This narrows the search space and speeds up tuning. However, some hyperparameters, such as $\alpha$ for \gls*{ffl} and the initial values for detector parameters, are selected randomly. In the future, we plan to develop a method for determining the distribution of frequencies. We can guess which frequency components are hard to synthesize, and choose a good $\alpha$ for \gls*{ffl} from the distribution of frequencies. We also plan to apply a \gls*{dl}-based approach to find a good initialization for detector parameters.

Our method of denoising is a physically-based Noise2Noise model, showing better results on our synthetic datasets than existing \gls*{dl}-based methods. On a real micrograph, with data augmentation using a state-of-the-art denoiser and neighboring projections, we show improvements compared to the state-of-the-art. For downstream applications, our denoised projections can contribute to better tomographic reconstruction and thus improve the quality of visualizations (\eg, \gls*{dvr}). In spite of the improvements, our denoising method shares the same limitation as in the original Noise2Noise~\cite{noise2noise}. This means that our method needs multiple projections to denoise one projection while in real-world scenario the data are limited. To mitigate this limitation, we believe it is possible to use neural networks to learn and transfer the patterns from various data. With the differentiability of our simulator, future work can be done to embed a neural network into the physics model of our simulator, taking advantage of the generalizability of neural networks while preserving physical intuitions.

\section*{Acknowledgment}

The authors would like to thank Sai Li and his team at School of Life Sciences, Tsinghua University, China for sharing the SARS-CoV-2 \gls*{cryo-em} data for this work.

\clearpage
\bibliographystyle{unsrt}
\bibliography{bibliography}

\clearpage
\section*{Supplementary Material}
\subsection*{More Comparisons of Denoising Synthetic Projections}
Here we present more comparisons of results of different synthetic settings to show the effectiveness and usefulness of our denoising method.

\begin{figure}[H]
\begin{center}
\includegraphics[trim={3.5cm 2cm 3.5cm 2cm},clip=true, width=\linewidth]{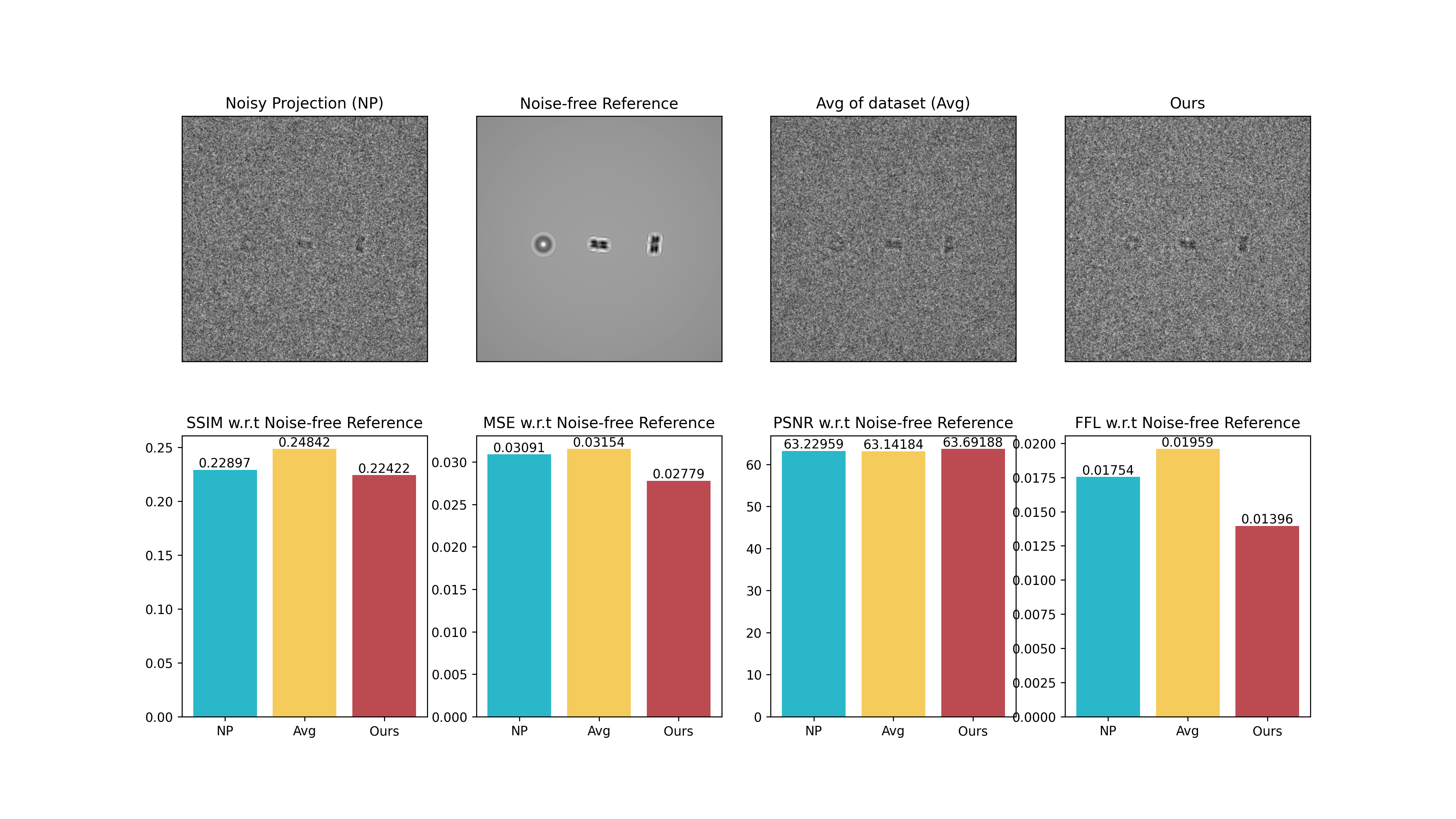}
\caption{Denoising visual and quantitative comparison of synthetic data with data setting (200, 2)}
\label{fig:num_2_dose_200}
\end{center}
\end{figure}

\begin{figure}[h]
\begin{center}
\includegraphics[trim={3.5cm 2cm 3.5cm 2cm},clip=true,width=\linewidth]{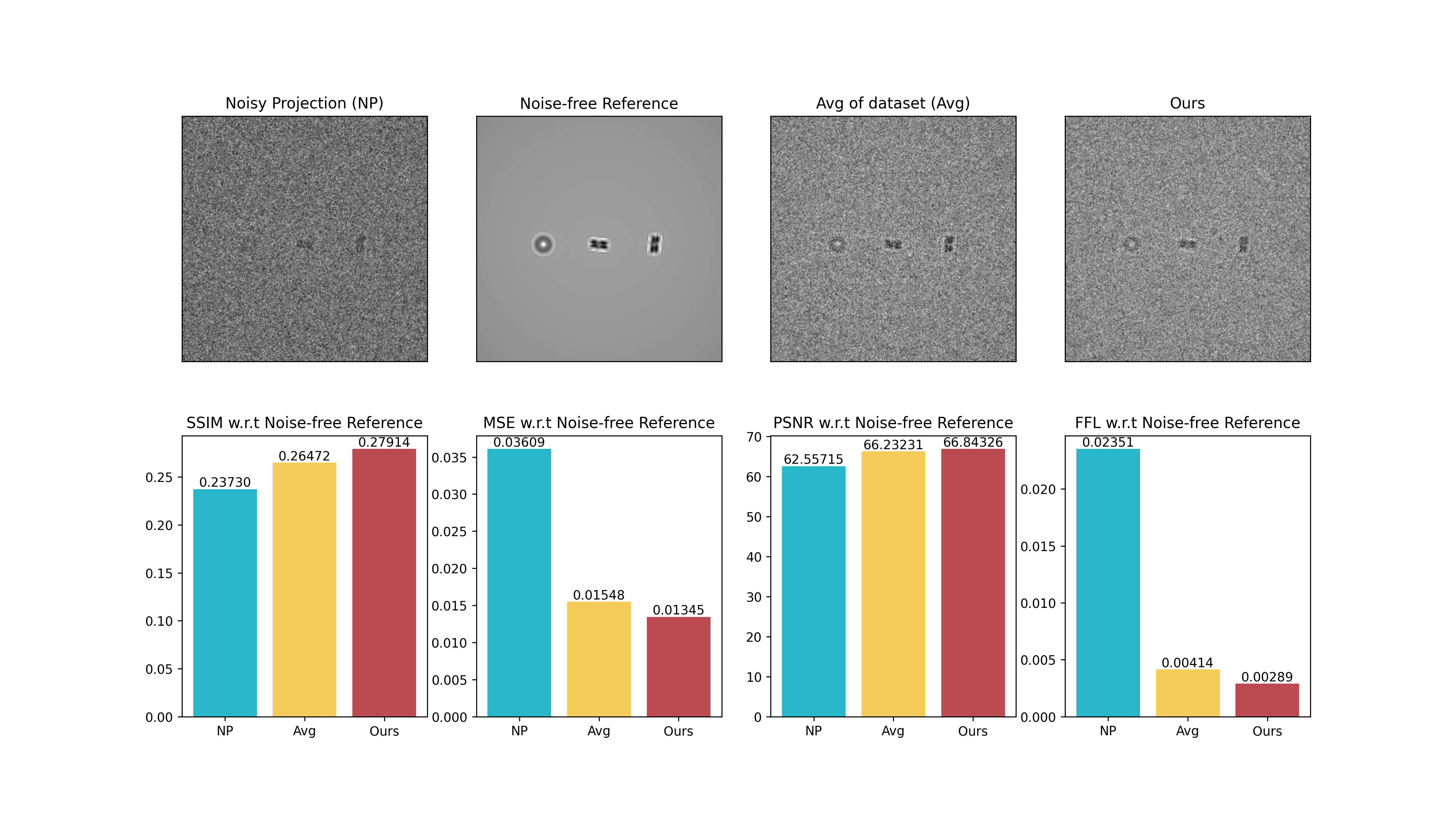}
\caption{Denoising visual and quantitative comparison of synthetic data with data setting (200, 5)}
\label{fig:num_5_dose_200}
\end{center}
\end{figure}

When the number of samples is small, for example, 2 or 5. Our results are close to the averages of the datasets visually. Yet, our results are better in terms of quantitative metrics as shown in \autoref{fig:num_2_dose_200} and \autoref{fig:num_5_dose_200} in which the results of data settings of (200, 2) and (200, 5) are presented.

\begin{figure}[H]
\begin{center}
\includegraphics[trim={3.5cm 2cm 3.5cm 2cm},clip=true,width=\linewidth]{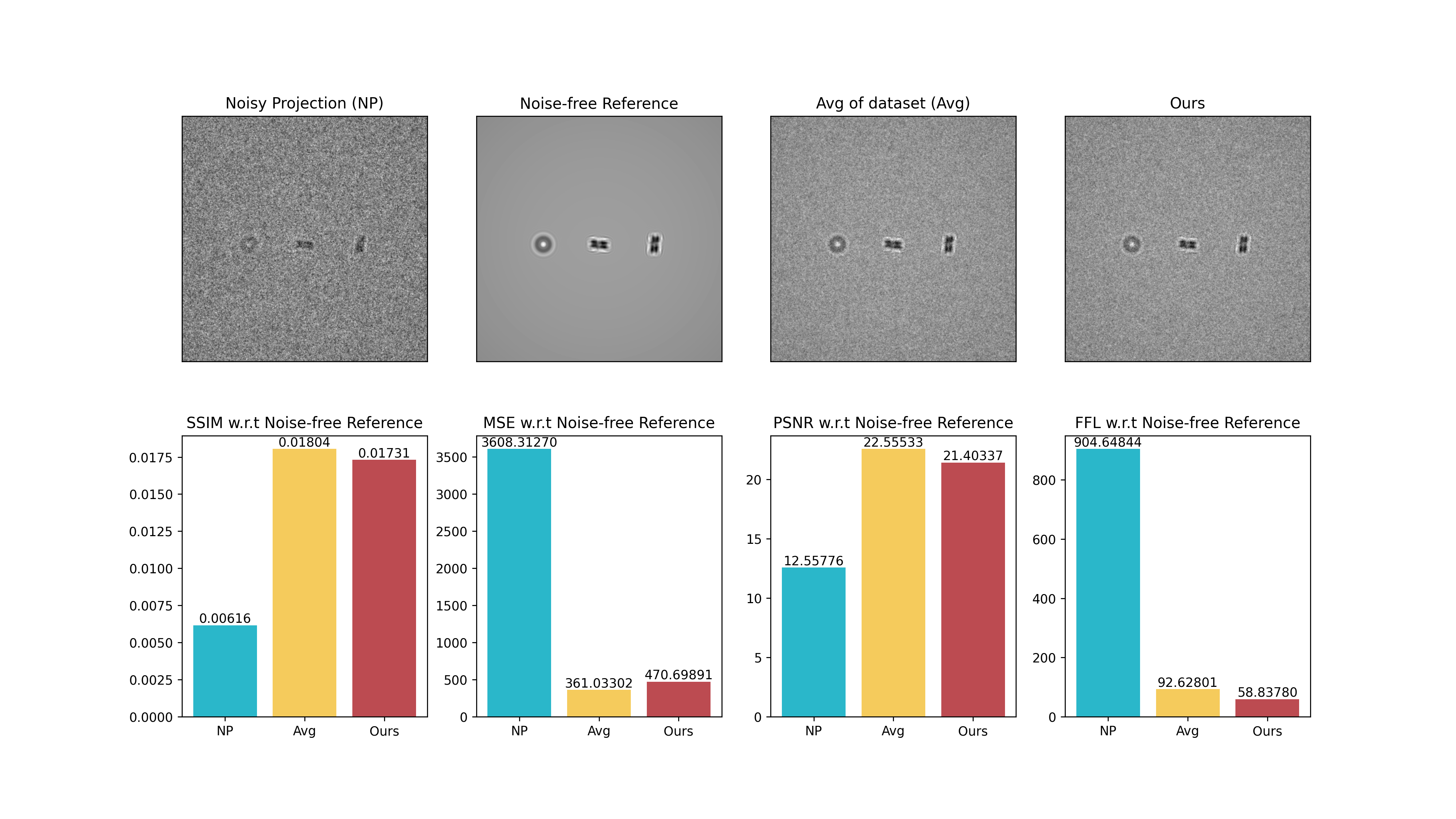}
\caption{Denoising visual and quantitative comparison of synthetic data with data setting (700, 10)}
\label{fig:num_10_dose_700}
\end{center}
\end{figure}

\begin{figure}[H]
\begin{center}
\includegraphics[trim={3.5cm 2cm 3.5cm 2cm},clip=true,width=\linewidth]{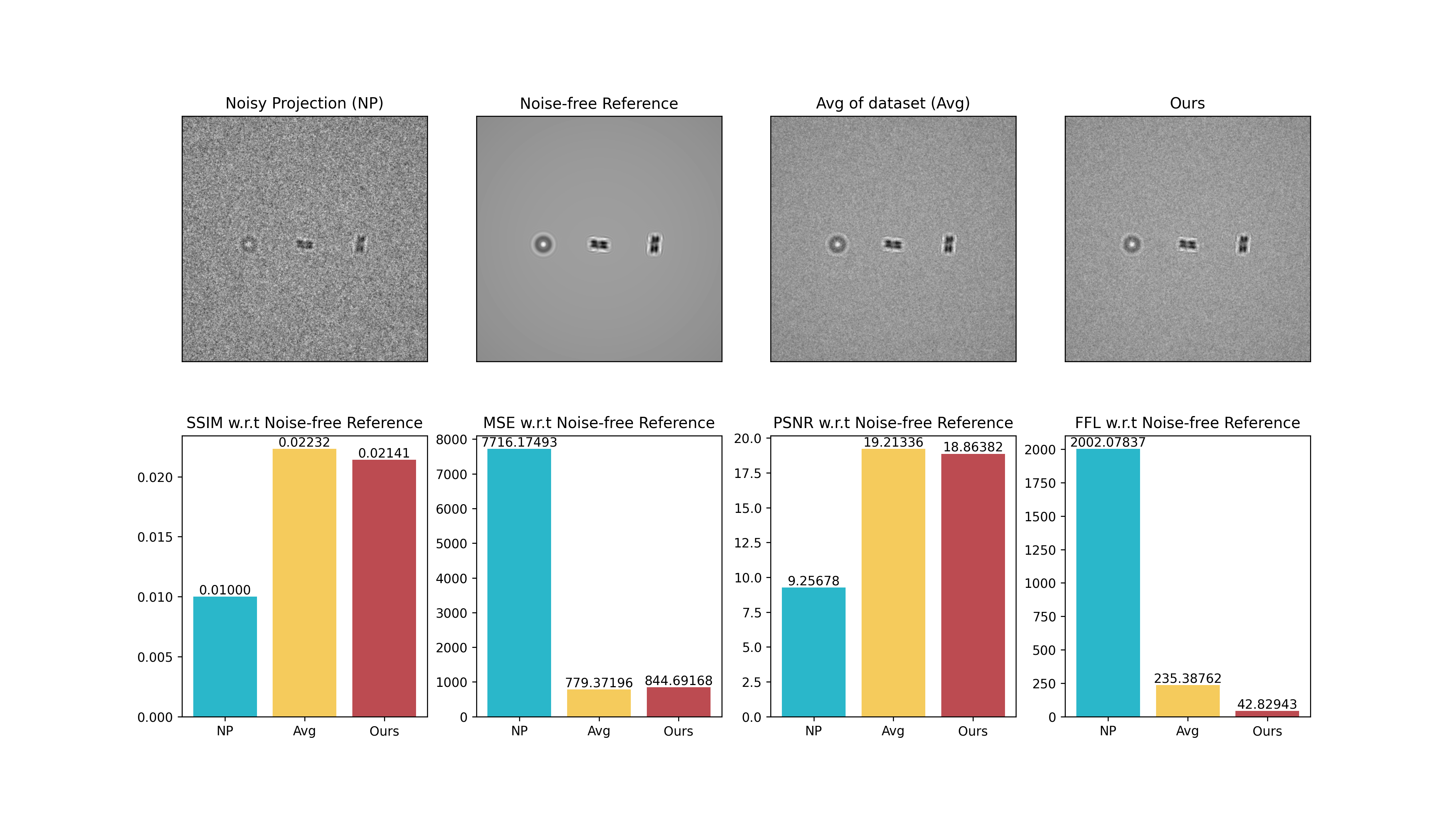}
\caption{Denoising visual and quantitative comparison of synthetic data with data setting (1500, 10)}
\label{fig:num_10_dose_1500}
\end{center}
\end{figure}

When the dose is high (\ie, greater than 700 electrons per nm\textsuperscript{2} per projection), our method consistently outperforms the averages of the datasets by a large margin in terms of \gls*{ffl}, which indicates our denoising method is much more correct in frequency domain. Here we present the results of the settings (700, 10) and (1500, 10). In \autoref{fig:num_10_dose_700} we can see the \gls*{ffl} is almost 50\% better than the average of the dataset and in \autoref{fig:num_10_dose_1500} the margin is more than 400\%.

Along with the results in \autoref{denoised_results_synthetic_setting}, we can conclude that despite the margins are different with different data settings, our method is advantageous over simple averaging consistently. This is especially beneficial when the data are too scarce and too noisy to train neural networks.

\end{document}